\documentclass[useAMS,usenatbib]{mn2e}
\usepackage{times}
\usepackage{amsmath}
\usepackage{amssymb}
\usepackage{longtable}
\usepackage{natbib}
\usepackage{epsfig}

\usepackage{adjustbox}
\usepackage{blindtext}

\title[Fast outflows in the metal-poor PN Hu\,1-2]
{Hu\,1-2: a metal-poor bipolar planetary nebula with fast collimated outflows}

\author[X.\ Fang et al.]{
X.\ Fang$^{1}$\thanks{E-mail: fangx@iaa.es}, 
M.~A.\ Guerrero$^{1}$, 
L.~F.\ Miranda$^{1}$, 
A.\ Riera$^{2}$, 
P.~F.\ Vel\'azquez$^{3}$, and 
A.~C.\ Raga$^{3}$ \\
$^{1}$Instituto de Astrof\'\i sica de Andaluc\'\i a (IAA, CSIC), 
Glorieta de la Astronom\'\i a s/n, E-18008 Granada, Spain\\
$^{2}$Departament de F\'\i sica i Enginyeria Nuclear, EUETIB, 
Universitat Polit\'{e}cnica de Catalunya, Comte d'Urgell 187, 
E-08036 Barcelona, Spain\\
$^{3}$Instituto de Ciencias Nucleares, Universidad Nacional Aut\'onoma de 
M\'exico, Apdo.\ Postal 70-543, CP: 04510, D. F., Mexico
}

\begin{document}

\date{Accepted . Received }

\pagerange{\pageref{firstpage}--\pageref{lastpage}} \pubyear{2014}

\maketitle

\label{firstpage}

\begin{abstract} 

We present narrow-band optical and near-IR imaging and optical long-slit 
spectroscopic observations of Hu\,1-2, a Galactic planetary nebula (PN) 
with a pair of [N~{\sc ii}]-bright, fast-moving ($>$340~km\,s$^{-1}$) 
bipolar knots. 
Intermediate-dispersion spectra are used to derive physical conditions 
and abundances across the nebula, and high-dispersion spectra to study 
the spatio-kinematical structure. 
Generally Hu\,1-2 has high He/H ($\approx$0.14) and N/O ratios 
($\approx$0.9), typical of Type~I PNe. 
On the other hand, its abundances of O, Ne, S, and Ar are low as compared 
with the average abundances of Galactic bulge and disc PNe.  
The position-velocity maps can be generally described as an hour-glass shaped 
nebula with bipolar expansion, although the morphology and kinematics of 
the innermost regions cannot be satisfactorily explained with a simple, 
tilted equatorial torus.  
The spatio-kinematical study confines the inclination angle of its major 
axis to be within 10$^{\circ}$ of the plane of sky.  
As in the irradiated bow-shocks of IC\,4634 and NGC\,7009, there 
is a clear stratification in the emission peaks of [O~{\sc iii}], 
H$\alpha$, and [N~{\sc ii}] in the northwest (NW) knot of Hu\,1-2.  
Fast collimated outflows in PNe exhibit higher excitation than other 
low-ionization structures.  
This is particularly the case for the bipolar knots of Hu1-2, with 
He~{\sc ii} emission levels above those of collimated outflows in 
other Galactic PNe. 
The excitation of the knots in Hu\,1-2 is consistent with the combined 
effects of shocks and UV radiation from the central star.  
The mechanical energy and luminosity of the knots are similar to those 
observed in the PNe known to harbor a post-common envelope (post-CE) 
close binary central star. 

\end{abstract}


\section{\label{intro} Introduction}

Planetary nebulae (PNe) represent the last stages in the evolution of 
low- and intermediate-mass stars, before they turn into white dwarfs. 
A large fraction of PNe presents complex morphologies, including 
axisymmetric shells, multipolar lobes, and collimated structures, in 
sharp contrast with the spherical envelopes typically seen around 
asymptotic giant branch (AGB) stars (e.g., Olofsson et al.\ 
\citealt{olo10}).  Somehow, the spherical AGB envelope is transformed 
into the complex PN shell. 
\emph{Hubble Space Telescope} (HST) observations have led to the 
suggestion that the shaping 
of the most structured young PNe (or proto-planetary nebulae, proto-PNe) 
could be attributed to the action of high-speed collimated outflows or 
jets that operate during the late-AGB and/or early post-AGB phase, as 
they interact with the intrinsically spherical AGB circumstellar envelope 
(Sahai \& Trauger \citealt{st98}).  These observations indicate that 
collimated outflows probably also play a crucial role in the shaping 
and dynamical evolution of PNe (e.g., Dennis et al.\ \citealt{den08}; 
Huarte-Espinosa et al.\ \citealt{hua12}). 

Collimated outflows or jet-like structures have been identified in a 
significant number of PNe.  The origin of collimated outflows is still 
uncertain, although they are most likely related to the evolution of 
binary systems, the action of magnetic fields, or both (e.g., Soker 
\citealt{sok06}; De~Marco \citealt{marco09}).  Particularly, binary 
interactions that involve accretion (e.g., Soker \citealt{sok98}; Soker
\& Livio \citealt{sl94}; Reyes-Ruiz \& L\'{o}pez \citealt{rrl99}; Nordhaus
\& Blackman \citealt{nb06}; Blackman \& Lucchini \citealt{bl14}) have been 
considered to be a plausible engine to produce the ubiquitous collimated 
outflows in PNe and proto-PNe (e.g., Balick \& Frank \citealt{bf02}).  
Recently, Tocknell, De~Marco \& Wardle \cite{toc14} constrained the 
physical properties of the common envelope (CE) interaction using the 
observed masses and kinematics of jets in four post-CE PNe. 

Detailed studies of the spectra of jet-like structures 
in PNe are scarce, but they generally report strong low-excitation 
emission lines (particularly from [N~{\sc ii}]) combined with 
high-excitation emission lines (e.g., [O~{\sc iii}]; Guerrero et al.\ 
\citealt{guerrero2008}; Gon\c{c}alves et al.\ \citealt{gon09}).  
This notable difference with the spectra of the main nebular shells can 
be attributed to the combined contributions of ionizing photons from the 
central star and excitation by shocks.  
Therefore, these structures are candidates of irradiated shocks, i.e., 
bow-shocks that are illuminated by ionizing (stellar) fluxes from the 
post-shock direction (e.g., Hartigan, Raymond \& Hartmann \citealt{har87}).

Hu\,1-2 (PN\,G086.5$-$08.8), first identified by Humason \cite{hum21} 
and classified as elliptical by Manchado et al. \cite{man96}, is a PN 
with fast, highly collimated outflows.  Recently, Miranda et al.\ 
\cite{mir12a} identified two bipolar compact knots along the main axis 
of Hu\,1-2 with bow-shock-like morphologies.  An analysis of the radial 
velocities and proper motions of these knots showed that they move at 
velocities $>$340 km~s$^{-1}$. 
This velocity is much higher than the expansion velocities of the 
collimated outflows observed in most PNe (Guerrero et al. \citealt{gue02}).  
For instance, the bright ``Saturn Nebula'', NGC\,7009, has an elliptical 
main nebula with a pair of [N~{\sc ii}]-bright outer knots along its 
major axis (e.g., Gon\c{c}alves et al.\ \citealt{gon03}).  
Despite the morphological similarities with Hu\,1-2, the expansion 
velocity of $\sim$60~km\,s$^{-1}$ reported for the bipolar knots of 
NGC\,7009 (Reay \& Atherton \citealt{ra85}) is much lower than that 
observed in Hu\,1-2, although we recall that the expansion velocity 
of the knots in NGC\,7009 is quite uncertain (Gon\c{c}alves et al.\ 
\citealt{gon03}). 

The high-velocity and bow-shock-like 
morphology led Miranda et al.\ \cite{mir12a} to conclude that these knots 
probably represent bow-shocks associated to high-velocity bullets. 
Furthermore, a preliminary analysis of narrow-band images of the knots 
suggests that [N~{\sc ii}] peaks farther away from the central star than 
[O~{\sc iii}] (Miranda et al.\ \citealt{mir12b}).  The enhanced [N~{\sc 
ii}] emission in the knots seems to imply shock-excitation, whereas the 
strong [O~{\sc iii}] emission rather points to irradiation from the 
central star.  A comprehensive analysis of the emission spectrum of the 
knots is crucial to determine the excitation mechanism, as done by, e.g., 
Riera et al.\ \cite{rie06} for Hen\,3-1475 and Guerrero et al.\ 
\cite{guerrero2008} for IC\,4634.

The physical structure of its main nebular shell is also largely 
unknown.  Narrow-band [N~{\sc ii}] image (Miranda et al.\ 
\citealt{mir12a}) reveals a complex morphology for its innermost 
regions, which adds to the peculiar velocity field revealed by 
kinematical studies (Sabbadin, Bianchini \& Hamzaoglu \citealt{sab83}; 
Sabbadin, Cappellaro \& Turatto \citealt{sab87}).  Similarly, the 
kinematical structure of the faint bipolar lobes is completely unknown.  
The relatively large He/H and N/O abundance ratios of the bright inner 
regions of Hu\,1-2 qualify it to be Type~I PN (Peimbert \& 
Torres-Peimbert \citealt{ptp87}), but detailed abundance analyses 
of the same nebular regions (Pottasch et al.\ \citealt{pot03}; Hyung, 
Pottasch \& Feibelman \citealt{hyu04}) found that abundances of the 
$\alpha$ elements (oxygen, neon, sulfur and argon) are much lower than 
those in most other Galactic PNe.  This peculiar abundance pattern may 
shed light on the origin and evolution of Hu\,1-2, but a spatially-resolved 
study of the physical conditions, chemical abundances, and excitation 
mechanisms is lacking.

In this paper we present high spatial resolution narrow-band optical 
and near-infrared images and high- and intermediate-dispersion long-slit 
optical spectra of Hu\,1-2.  These data have allowed us to carry out a 
complete analysis of its spatio-kinematical properties and physical 
conditions, abundances, and excitation mechanisms.  Particular emphasis 
is made in the investigation of the properties of the collimated outflows 
in Hu\,1-2.

\begin{figure*}
\begin{center}
\epsfig{file=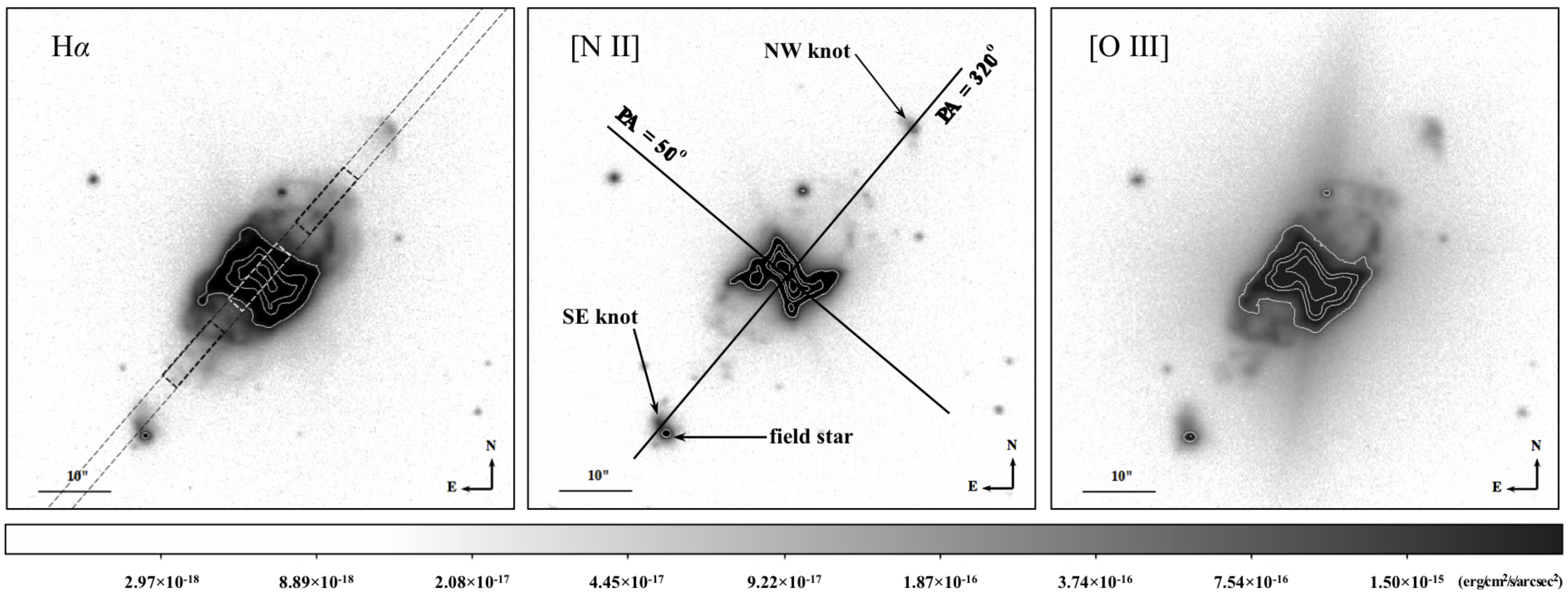,width=17.5cm,angle=0}
\caption{
Nordic Optical Telescope (NOT) ALFOSC flux-calibrated grey-scale
images of Hu\,1-2\ in H$\alpha$ (left), [N~{\sc ii}] (middle)
and [O~{\sc iii}] (right).  The 2\farcs5-wide long slit along PA
$\approx$320$^{\circ}$ used for the intermediate-dispersion spectroscopy
(see Section~\ref{observe:2}) was placed across the northwest (NW) and 
southeast (SE) knots as
indicated by long-dashed lines in the H$\alpha$ image, where the slit
apertures used to extract 1-D spectra for the inner and outer regions
are indicated by the heavy white and black dashed boxes (10\arcsec\ each
in length), respectively.  The NW and SE knots, as well as a field star
partially superimposed on the SE knot, are indicated in the [N~{\sc ii}]
image, where the two slit positions used for high-dispersion spectroscopy
(see Section~\ref{observe:3}) are indicated with two black lines (slit
width not to scale).  The ``spikes'' along the north-south direction in
the [O~{\sc iii}] image are stray light due to the very strong [O~{\sc
iii}] emission from the inner region of Hu\,1-2.  Images are scaled to
the same level and the grey levels are in logarithm.  The white contours
over-plotted in the central regions correspond to 2.5$\times$10$^{-14}$,
1.0$\times$10$^{-14}$, 5.0$\times$10$^{-15}$, and 1.0$\times$10$^{-15}$
erg\,cm$^{-2}$\,s$^{-1}$\,arcsec$^{-2}$.  The color bar below shows the
grey scale in surface brightness
(in erg\,cm$^{-2}$\,s$^{-1}$\,arcsec$^{-2}$). }
\label{grey_img}
\end{center}
\end{figure*}

\section{\label{observe} Observations}

\subsection{\label{observe:1} Imagery}

Narrow-band optical images in the [O~{\sc iii}], H$\alpha$, and [N~{\sc ii}] 
emission lines were acquired on 2008 September 2 using the Andaluc\'{i}a 
Faint Object Spectrograph and Camera (ALFOSC) on the 2.5m Nordic Optical 
Telescope (NOT) of the Observatorio del Roque de los Muchachos (ORM) on 
the island of La Palma (Spain).  The characteristics of the narrow-band 
filters used in these observations (central wavelength $\lambda_{\rm c}$ 
and bandwidth $\Delta\lambda$) are summarized in Table~\ref{imagery}. 
An EEV 2k$\times$2k CCD was used as detector, yielding a plate 
scale of 0\farcs184 pixel$^{-1}$ and a field of view (FoV) of 
6\farcm3$\times$6\farcm3.

Two images in each filter were obtained with a small dithering between 
them to eliminate cosmic rays and reduce cosmetic defects of the CCD. 
The images were bias-subtracted, flat-fielded by twilight flats, and 
combined using standard {\sc iraf}\footnote{{\sc iraf}, the Image 
Reduction and Analysis Facility, is distributed by the National Optical 
Astronomy Observatory, which is operated by the Association of 
Universities for Research in Astronomy under cooperative agreement with 
the National Science Foundation.} V2.14.1 routines.  The spatial resolution 
of the final images, as derived from the FWHM of stars in FoV is 0\farcs7.

\begin{figure}
\begin{center}
\epsfig{file=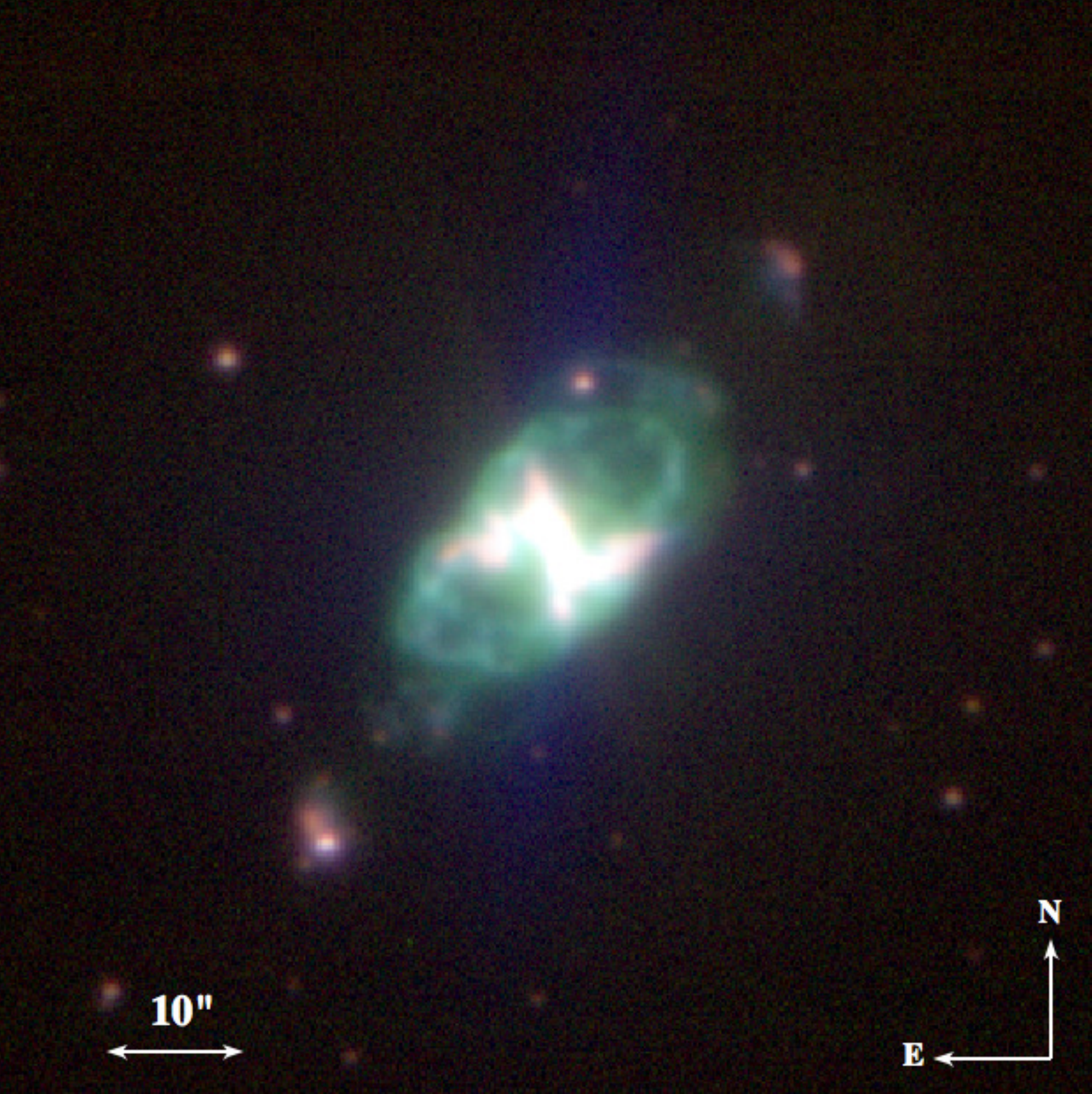,width=8.3cm,angle=0}
\caption{Color-composite picture of Hu\,1-2 in the [N~{\sc ii}] (red), 
H$\alpha$ (green), and [O~{\sc iii}] (blue) emission lines.  All intensities 
are displayed in logarithmic scale.  A field star is partially superimposed 
on the southeast (SE) knot. }
\label{color_img}
\end{center}
\end{figure}

\begin{figure}
\begin{center}
\epsfig{file=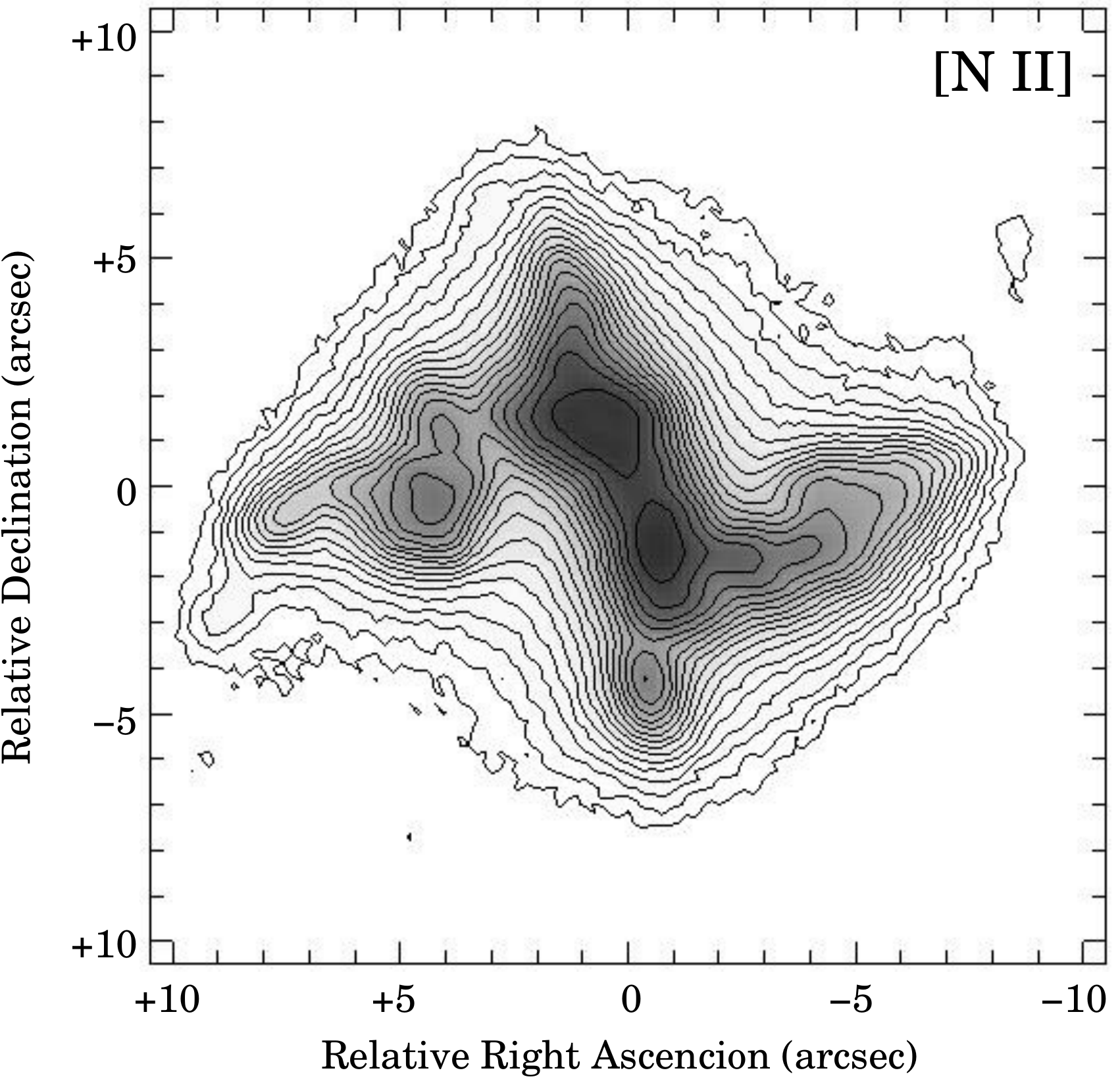,width=8.0cm,angle=0}
\caption{Grey-scale NOT ALFOSC [N~{\sc ii}] narrow-band image of Hu\,1-2 
over-plotted with contours, showing the complex knotty structure in its 
central regions.  The image is centered on the central star (as can be 
seen in Figure~\ref{ir_img}).  Grey scale is in logarithm and the contour 
levels are arbitrary. }
\label{nii_cont}
\end{center}
\end{figure}

The three images were then background-subtracted and flux-calibrated using 
the H$\alpha$, [N~{\sc ii}] $\lambda$6583 and [O~{\sc iii}] $\lambda$5007 
line fluxes measured in the intermediate-dispersion spectrum that will be 
presented in Section~2.2.  The individual flux-calibrated narrow-band 
images of Hu\,1-2 are presented in Figure~\ref{grey_img}, whereas a 
color-composite picture is displayed in Figure~\ref{color_img}.  In 
Figure~\ref{nii_cont} we present a close-up of the [N~{\sc ii}] emission 
from the inner region of Hu\,1-2.

Narrow-band, near-infrared (IR) images were obtained on 2004 July 11 with 
the 3.5m Telescope Nazionale Galileo (TNG) also at the ORM.  The 
Near-Infrared Camera and Spectrograph (NICS) was used with a 
1024$\times$1024 Rockwell HgCdTe array.  The spatial scale is 0\farcs13 
pixel$^{-1}$ and the FoV is 2\farcm2$\times$2\farcm2.  The images were 
obtained through Br$\gamma$, H$_2$ and continuum ($K_{\rm c}$) filters 
whose central wavelengths and bandwidths are given in Table~\ref{imagery}. 
The images were reduced with the {\sc midas}\footnote{{\sc midas} is 
developed and distributed by the European Southern Observatory.} package 
following standard procedures for near-IR image reduction.  The spatial 
resolution of the final images, as derived from the FWHM of field stars 
in the FoV, is 0\farcs75.  Figure~\ref{ir_img} presents the three near-IR 
images.

\begin{table}
\begin{center}
\caption{Observing log of imaging.}
\label{imagery}
\begin{tabular}{llccrr}
\hline
\multicolumn{1}{l}{Telescope}  & 
\multicolumn{1}{l}{Instrument} & 
\multicolumn{1}{c}{Filter}     & 
\multicolumn{1}{c}{$\lambda_{\rm c}$} & 
\multicolumn{1}{c}{$\Delta\lambda$} & 
\multicolumn{1}{c}{Exp.\ Time} \\ 
\multicolumn{1}{l}{} & 
\multicolumn{1}{l}{} & 
\multicolumn{1}{c}{} & 
\multicolumn{1}{c}{} & 
\multicolumn{1}{c}{} & 
\multicolumn{1}{c}{(s)} \\
\hline
NOT & ALFOSC & [O~{\sc iii}] & 5007\,{\AA} & 30\,{\AA}~~~ & 600~~~ \\
    &        & H$\alpha$     & 6567\,{\AA} &  8\,{\AA}~~~ & 600~~~ \\
    &        & [N~{\sc ii}]  & 6588\,{\AA} &  9\,{\AA}~~~ & 900~~~ \\
\hline
TNG & NICS   & Br$\gamma$    & 2.169 $\mu$m & 0.035 $\mu$m & 1200~~~ \\
    &        & H$_{2}$        & 2.122 $\mu$m & 0.032 $\mu$m & 1200~~~ \\
    &        & $K_{\rm c}$    & 2.275 $\mu$m & 0.039 $\mu$m & 1200~~~ \\
\hline
\end{tabular}
\end{center}
\end{table}

\subsection{\label{observe:2} Intermediate-dispersion spectroscopy}

Intermediate-dispersion spectra of Hu\,1-2 were obtained on 2011 October 
9, using the ALBIREO spectrograph at the 1.5m telescope of the 
Observatorio de Sierra Nevada (OSN), Granada, Spain.  A Marconi 
2048$\times$2048 CCD was used as a detector, in conjunction with the 400 
lines~mm$^{-1}$ grating \#4 blazed at 5500\,{\AA}.  The slit length was 
$\sim$ 6\arcmin\ and its width was set at 50~$\mu$m ($\equiv$2\farcs5).  
A 2$\times$2 binning in the detector was used, implying plate and spectral 
scales of 1\farcs53 pixel$^{-1}$ and 3.54\,{\AA}~pixel$^{-1}$, respectively. 
The spectral resolution was $\sim$4.7\,{\AA}, and the wavelength 
uncertainty $\sim$1\,{\AA}. The spectral coverage is 3570--7200\,{\AA}.
The seeing, as determined from the FWHM of the continuum of field stars
covered by the slit, was $\sim$3\arcsec. As illustrated in 
Figure~\ref{grey_img}-\emph{left}, the slit was aligned along a position 
angle (PA) of 320\degr, i.e., along the major nebular axis and through 
the bipolar knots.

Three 60\,s and six 300\,s exposures were obtained to secure information
both from the bright and faint emission lines in this nebula;  indeed,
the bright [O~{\sc iii}] $\lambda$5007 emission line was found to
saturate in the central regions of the nebula in the long exposures.
All the spectra were bias-subtracted, flat-fielded, wavelength and
flux calibrated, following standard procedures using {\sc iraf}.
Spectra of the spectrophotometric standard star Feige~115 acquired on the
same night were used to carry out the flux calibration.

\subsection{\label{observe:3} High-dispersion spectroscopy}

Two high-dispersion, long-slit spectra of Hu\,1-2 were obtained in 
2004 June with IACUB\footnote{
The IACUB uncrossed echelle spectrograph was built in a collaboration 
between the Instituto de Astrof\'{i}sica de Canarias (IAC) and Queen's 
University of Belfast.} on the 2.5m NOT at ORM.  The slit positions, 
oriented at PAs 50$^{\circ}$ and 320$^{\circ}$, are shown in 
Figure~\ref{grey_img}-\emph{middle}.  The long-slit spectra at PA 
320$^{\circ}$ was already presented by Miranda et al. \cite{mir12a} in 
the analysis of the kinematics of the bipolar knots. 
The H$\alpha$ and [N~{\sc ii}] $\lambda$$\lambda$6548,\,6583 emission 
lines were observed with a slit width of 0\farcs65 for an exposure 
of 900\,s.  The spectral resolution (FWHM) is 8\,km\,s$^{-1}$, and the 
spatial resolution is $\simeq$\,1\arcsec.
See Miranda et al. \cite{mir12a} for a detailed description of these 
observations.


\section{Main nebula}

\subsection{Morphology}

The optical narrow-band images in Figure~\ref{grey_img} show that Hu\,1-2
has an elongated elliptical or slightly bipolar main nebula with a size 
of $\sim$12\arcsec$\times$32\arcsec\ with the major axis oriented at
PA$\sim$320$^{\circ}$, and a pair of bipolar knots (NW and SE) located 
along the main nebular axis, each $\sim$27\arcsec\ away from the central 
star.  As already noted by Miranda et al. \cite{mir12a}, the SE knot is 
partially superimposed by a field star and cannot be well studied. At any
rate, the high-resolution NOT images reveal that the knots (in particular 
the NW one) present bow-shock morphology (Figure~\ref{grey_img}). In the 
central region of Hu\,1-2, there is a bright, $z$-shaped structure 
(hereafter the inner region) with a size of 
$\sim$10\arcsec$\times$10\arcsec\ (Figure~\ref{grey_img}-\emph{middle}; 
see also description below). The bipolar lobes are considerably fainter 
than the inner region. Moreover, the lobes harbor a noticeable richness 
of small structures that differ from each other. The NW lobe seems to be 
composed of two concentric structures, a NW inner lobe and a NW outer 
lobe, that are better observed in the H$\alpha$ image.  Only part of the 
NW outer lobe can be recognized in the [O~{\sc iii}] image, where it shows 
a clearly curly morphology, and in the [N~{\sc ii}] image, where the 
[N~{\sc ii}] emission is mainly detected in a couple of very compact 
regions.  A SE counterpart of the NW inner lobe can be recognized, also
exhibiting a remarkable curly structure in the [O~{\sc iii}] image. No SE 
counterpart of the NW outer lobe can be identified.  Instead, several 
knots are observed in the [O~{\sc iii}] image, that are oriented along 
(or close to) the main nebular axis.  Faint [N~{\sc ii}] emission is 
detected at the leading head of these knots.

\begin{figure*}
\begin{center}
\epsfig{file=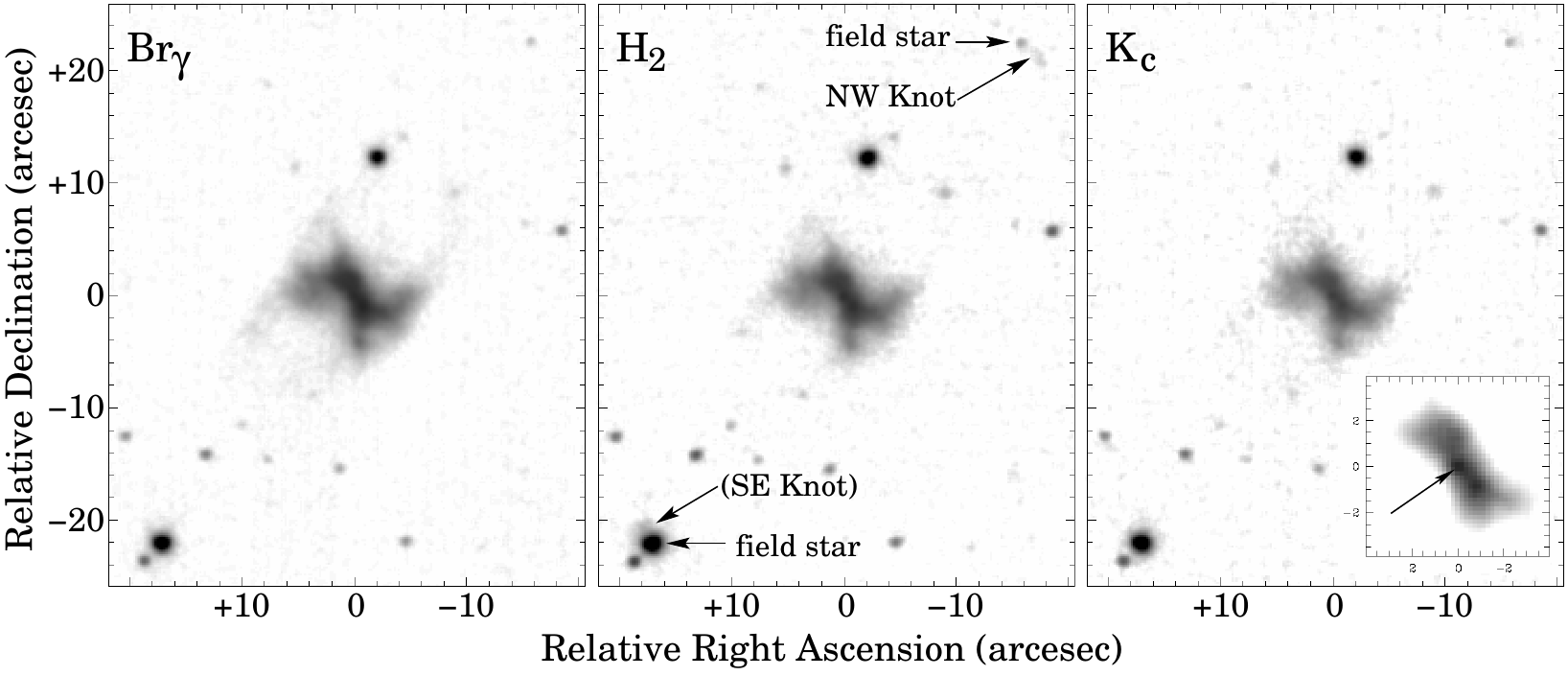,width=17.5cm,angle=0}
\caption{Grey-scale reproductions of the narrow-band, near-IR images of
Hu\,1-2.  Grey levels are logarithmic  and have been chosen to emphasize
the main nebular structures detected.  The NW knot, faint emission possibly
from the SE knot, and two field stars are indicated.  The inset in the
$K_{\rm c}$ image (right panel) shows the inner 4\arcsec$\times$4\arcsec\
central region of Hu\,1-2 in linear grey levels. The arrow indicates a
point-like source at the nebular's center that may correspond to the
central star. }
\label{ir_img}
\end{center}
\end{figure*}

A close inspection of the inner region of Hu\,1-2 in the [N~{\sc ii}] image 
reveals a peculiar structure, as shown in Figure~\ref{nii_cont}.  It shows 
an arc-shaped central ``bar'' with a size $\simeq$6\arcsec\ oriented 
at PA $\sim$30$^{\circ}$ that contains two bright knots separated by 
$\simeq$2\farcs8.  There is enhanced emission and/or knots on either end 
of the bar that seem to trace the edges of bipolar lobes towards the east 
and west.  As a whole, the inner region of Hu\,1-2 exhibits a remarkable 
point-symmetric structure.  It is worth noting that none of the 
point-symmetric pairs of knots is oriented along the major nebular axis.  
It is tantalizing to interpret the inner region as a broken equatorial 
torus, but we have to admit that such a torus would be broken in a very 
point-symmetric manner.  The kinematics of this region, to be described in 
Section~3.2, is less compatible with that expected from a simple ring-like 
equatorial torus.

The optical images (Figures~\ref{grey_img} and \ref{color_img}) show that 
[N~{\sc ii}] emission is enhanced in the knots and in the inner region 
while the bipolar lobes (hereafter the outer region) are dominated by the 
H$\alpha$ line emission.  The strongest [O~{\sc iii}] emission comes from 
both the main nebula and regions behind the head of the knots facing 
toward the central star.

The point-symmetric inner region is also enhanced in the near-IR emission, 
as shown in Figure~\ref{ir_img}, where it presents a very similar morphology
to that observed in the optical ones.  In addition, faint emission from the 
bipolar lobes can be recognized in the Br$\gamma$ filter.  The H$_{2}$ 
emission is faint, but definitely present in the NW knot and, possibly, also 
in the SE counterpart.  We have tried to subtract the continuum image from 
the emission-line images (the subtracted images are not shown here), and 
found that Br$\gamma$ emission is present in the inner region and bipolar 
lobes whereas in the case of H$_{2}$, no satisfactory subtraction was 
achieved; nevertheless, the presence of H$_{2}$ emission in the inner 
region is doubtful as the emission in the H$_{2}$ image can be attributed 
to continuum emission.  Finally we note that both the $K_{\rm c}$ and 
H$_{2}$ images show a faint point-like source at the center of the inner 
region (Figure~\ref{ir_img}), which could be the central star of Hu\,1-2. 
This point-like source is not well observed in the Br$\gamma$ image 
probably due to the relatively stronger line emission.  In all three 
near-IR filters, we observed an emission feature located towards the 
SW of the central point-like source (better seen in 
Figure~\ref{ir_img}-\emph{right}). 
A comparison between the NOT ALFOSC [N~{\sc ii}] and TNG NICS 
$K_{\rm c}$ images shows that the SW feature in the $K_{\rm c}$ image 
is associated with the SW [N~{\sc ii}] knot in the central bar.  The 
$K_{\rm c}$ knot peaks closer to the central star by $\sim$0\farcs3, 
suggesting that it corresponds to ionized material with higher 
excitation than the [N~{\sc ii}] line.

\begin{figure}
\begin{center}
\epsfig{file=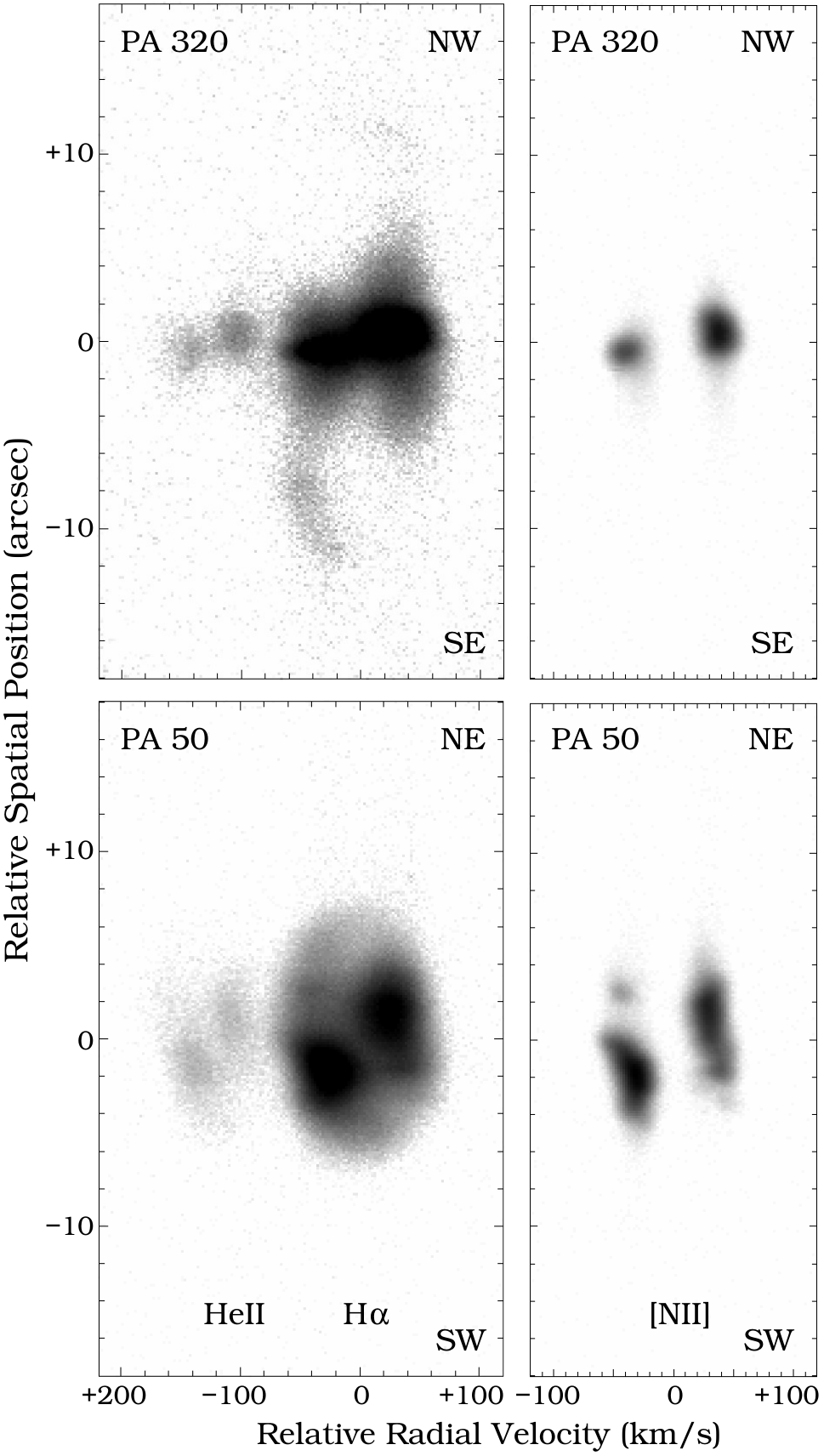,width=7.5cm,angle=0}
\caption{Grey-scale PV maps of the He~{\sc ii} and H$\alpha$ lines (left) 
and the [N~{\sc ii}] line (right). 
In the left panels, the two weak features bluewards of H$\alpha$ 
belong to the He~{\sc ii} $\lambda$6560 line.  The slit PA = 
320$^{\circ}$ and 50$^{\circ}$ (see Figure~\ref{grey_img}-\emph{middle}) 
are shown in the upper and lower panels, respectively. }
\label{kinematics}
\end{center}
\end{figure}

\subsection{Kinematics}

The grey-scale position-velocity (PV) maps of the H$\alpha$ and 
[N~{\sc ii}] $\lambda$6583 emission lines derived from the 
high-resolution, long-slit spectra at PA 50$^{\circ}$ and 320$^{\circ}$ 
are shown in Figure~\ref{kinematics}.  The PV map of the He~{\sc ii} 
$\lambda$6560 emission line is close to that of H$\alpha$. 
The PV map of the [N~{\sc ii}] line at PA 320$^{\circ}$ was already 
presented by Miranda et al.\ \cite[][Figure\,2 therein]{mir12a}, who 
focused on the emission features associated with the bipolar knots.  
We examine in detail the kinematical structure of the main nebula 
instead.  
In the following, radial velocities will be quoted with respect 
to the heliocentric systemic velocity of $-$3.3~km\,s$^{-1}$ for 
Hu\,1-2 that we deduced from our high-dispersion spectra (see below).

The [N~{\sc ii}] PV map at PA=320$^{\circ}$ shows two compact features at 
$\simeq$0\farcs5 from the center with radial velocities $\pm$38 km~s$^{-1}$, 
with the feature at PA 320$^{\circ}$ being redshifted.  
In the PV map at PA 50$^{\circ}$, the emission features are more elongated
along the slit ($\simeq$10\arcsec), and present point-symmetric, arc-like
shapes, with the NE feature redshifted and the SW one blueshifted.  
The emission peaks in these features are located $\simeq$1\farcs9 from 
the center and their radial velocity amounts to $\pm$30 km~s$^{-1}$.  
At larger distances from the center, the radial
velocity in the features remains approximately constant; at smaller 
distances the radial velocity increases up to $\simeq$ $-$48 
km\,s$^{-1}$ at $\simeq$ 0\farcs3 from the center in the blueshifted 
feature and up to $\simeq\,+$43 km\,s$^{-1}$ at 1\farcs5--3\farcs2 
from the center in the redshifted feature.  
This morphology in the PV diagram indicats a rotation of the torus. 
If the inner structure is/was indeed a ring (or torus), then it could 
have been distorted by some agent such as several bipolar ejections 
along different directions, as suggested by the point-symmetry of the 
bright regions.  If this is the case, the kinematics cannot be 
interpreted easily.  An isolated knot is also observed at $-$43 
km\,s$^{-1}$ and 2\farcs5 (Figure~\ref{kinematics}), which reflects 
the complexity of the inner structure.

These [N~{\sc ii}] features are generally consistent with the brightest 
features in the H$\alpha$ PV maps.  The arc shape of the [N~{\sc ii}] 
features in the PV map at 
PA 50$^{\circ}$ can also be recognized in H$\alpha$ with radial 
velocities about 5~km\,s$^{-1}$ lower than those measured in the 
[N~{\sc ii}] emission line.  
In the H$\alpha$ PV map, these features are embedded or superimposed on 
a broad and faint H$\alpha$ component that extends to $\simeq$12\arcsec\ 
and $\simeq$130~km\,s$^{-1}$ in radial velocity.  
The broad H$\alpha$ emission is mostly symmetrical with 
respect to the velocity axes.  At PA 320$^{\circ}$ the extended H$\alpha$ 
emission is observed up to a distance of 12\arcsec\ on both sides of the 
center.  This emission corresponds to the elliptical/bipolar lobes of the 
nebula.  Although the emission is very faint in the PV maps, the observed 
kinematics appears compatible with that expected from an hour-glass nebula. 
The total spatial extent of the H$\alpha$ emission detected in 
the PV maps (12\arcsec$\times$24\arcsec) coincides very well with 
the spatial extent of the main nebula observed in direct images.

Two bright features are detected in the He~{\sc ii} emission line 
at both PAs.  Their radial velocity separation amounts to $\simeq$
33\,km\,s$^{-1}$, while their spatial separation is $\simeq$1\farcs7 
at PA 50$^{\circ}$ (where they present a slight spatial elongation) and
$\simeq$0\farcs4 at PA 320$^{\circ}$ (Figure~\ref{kinematics}).  
The brightest features in each of the three emission lines and their 
spatio-kinematical properties indicate that they probably trace a 
unique structure that is related to the bright inner region of Hu\,1-2.

It has been shown that differing the systemic velocities ($v_{\rm sys}$) 
of the main nebular shell and collimated outflows can be associated to 
the presence of a binary central star (e.g., Miranda et al.\ \citealt{mir01}). 
Using our high-dispersion spectra of Hu\,1-2, we measured the heliocentric 
systemic velocities of the main nebula and the knots: 
$v_{\rm sys}$ (nebula) = $-$3.3$\pm$1.7 km\,s$^{-1}$, and 
$v_{\rm sys}$ (knot)   = $+$1.4$\pm$2.5 km\,s$^{-1}$.  
These two systemic velocities agree within the errors and, therefore, 
they cannot confirm whether a binary central star exist or not.  
We note that our measurement of the nebular systemic velocity are 
consistent, but not completely coincident, with previous measurements: 
$-$9.0$\pm$8.1~km\,s$^{-1}$ (Schneider et al.\ \citealt{sch83}) and 
$+$9.0$\pm$4.3~km\,s$^{-1}$ (Durand et al.\ \citealt{dur98}). 
These differences most likely reflect the complex kinematical structure 
of the main nebular shell of Hu\,1-2, which makes difficult a precise 
estimate of the systemic velocity based on spatially unresolved 
observations.

\subsection{Spatio-kinematical model}

\begin{figure}
\begin{center}
\epsfig{file=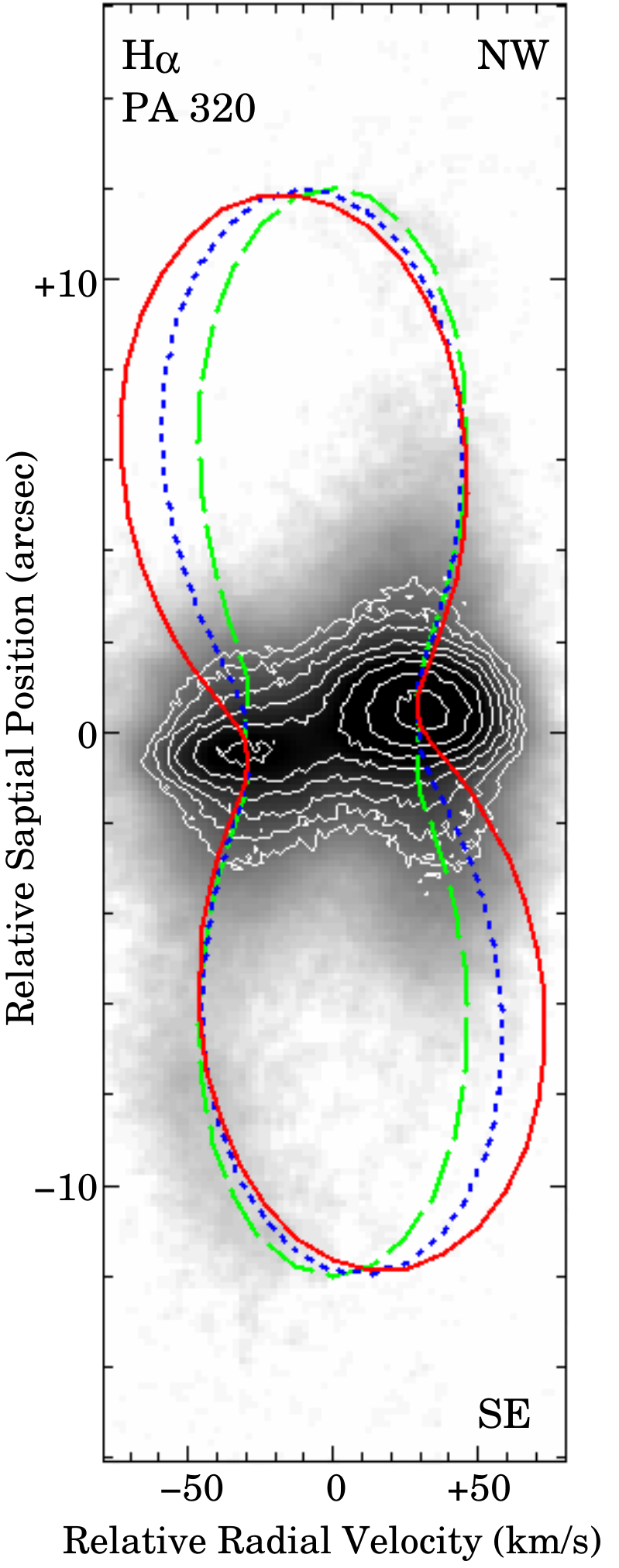,width=6.5cm,angle=0}
\caption{
Grey-scale PV map of the H$\alpha$ emission line at PA = 320$^{\circ}$. 
A smoothing with a 3$\times$3 pixel box has been applied to the 
original PV map (see also Figure~\ref{kinematics}).  The over-plotted white 
contours define the position of the two emission maxima.  Also over-plotted 
are three hour-glass models ($v_{\rm p}$ = 150\,km\,s$^{-1}$, $R_{\rm p}$ 
= 12\farcs5) with different inclination angles of the polar axis with 
respect to the plane of the sky, $\theta$, and the model parameter 
$\gamma$ (see Equation~\ref{eq1}):
$\theta$ = 10$^{\circ}$ and $\gamma$ = 2.5 (red solid line), 
$\theta$ = 5$^{\circ}$ and $\gamma$ = 3.5 (blue dotted line), and 
$\theta$ = 0$^{\circ}$ and $\gamma$ = 4.5 (green dashed line). }
\label{model}
\end{center}
\end{figure}

The H$\alpha$ PV map at PA 320$^{\circ}$ is similar to that found in 
other bipolar nebula (e.g., Kn\,26; Guerrero et al. \citealt{guerrero2013}) 
suggesting bipolar expansion.  
We have adopted the prescription by Solf \& Ulrich \cite{su1985} to model 
a bipolar outflow, where the radial velocity ($v_{\rm r}$) at a latitude 
angle ($\phi$) above the equatorial plane is given by
\begin{equation}
\label{eq1}
v_{\rm r}(\phi) = v_{\rm p} + (v_{\rm p} - v_{\rm e}) \times
sin^{\gamma}(\phi)
\end{equation}
where $v_{\rm p}$ and $v_{\rm e}$ are the polar and equatorial velocities,
respectively, and $\gamma$ is a parameter to fit the specific shape of the 
hour-glass.  We note that the PV map suggests a small inclination angle 
with respect to the plane of the sky. 
Figure~\ref{model} shows three hour-glass models at inclination angles of 
10$^{\circ}$, 5$^{\circ}$, and 0$^{\circ}$, each assuming that $v_{\rm p}$ 
and $v_{\rm e}$ are 150 and 30\,km\,s$^{-1}$, respectively, and a polar 
radius ($R_{\rm p}$) of 12\farcs5.  By varying the parameter $\gamma$ we 
obtained the best fit model for the PV map.

It is clear from Figure~\ref{model} that no ideal representation of the 
kinematics can be reached.  
Whereas one particular set of parameters explains the observed kinematics 
of a given region (e.g., the front side of the SE lobe 
and the rear side of the NW lobe), other parameters are needed to reproduce 
the observed kinematics of other nebular regions (e.g., the rear side of 
the SE lobe and the front side of the NW lobe).  This is not surprising 
because the images already suggest noticeable morphological differences 
between the two lobes, and these are reflected into differences in their 
intrinsic kinematics.  From our analysis we conclude that a reasonable 
upper limit for the inclination angle of the polar axis is 10$^{\circ}$ 
and an approximate value for the polar expansion velocity is 
150\,km\,s$^{-1}$.  Assuming a lower limit of 3.5~kpc for the distance 
to Hu\,1-2 (Miranda et al.\ \citealt{mir12a}), we derived a lower limit 
$\simeq$1100\,yr for the kinematical age of the bipolar lobes, which is 
consistent with the value of 1375$^{+590}_{-320}$ yr obtained for the 
bipolar knots (Miranda et al.\ \citealt{mir12a}).

The [N~{\sc ii}] features in the PV maps could be due to a tilted torus 
(or ring-like structure).  If we assume a diameter of 10\arcsec\ for the 
equator and a projection of 1\arcsec\ due to tilt, an inclination angle 
$\simeq$6$^{\circ}$ is obtained for the plane of the ring with respect to 
the line of sight.  This value is compatible with the inclination angle of 
the bipolar lobes.  However, the observed kinematics strengthens the idea 
that the inner regions of Hu\,1-2 is difficult to be interpreted as a 
simple equatorial torus.  In particular, for a torus oriented at PA 
$\simeq$50$^{\circ}$ and seen almost edge on, a long-slit spectrum along 
PA $\simeq$50$^{\circ}$ should show a velocity ellipse with maximum 
splitting at the center and decreasing towards the edges of the torus. The 
PV map at PA $\simeq$50$^{\circ}$ (Figure~\ref{kinematics}) shows a very 
different structure.  A possible interpretation for the inner region is 
that a series of bipolar ejections at different directions have have been 
involved in its formation, distorting a previous, more defined structure.

\begin{table*}
\caption{
Emission lines detected in the inner region, outer region, and NW knot 
of Hu\,1-2.  The observed fluxes and the extinction corrected intensities 
are all normalized to H$\beta$ = 100.  The colon ``:'' indicates a large 
uncertainty in the line intensity. 
Logarithm of the total observed H$\beta$ fluxes measured from the 
1-D spectra extracted from the slit apertures shown in 
Figure~\ref{grey_img}-\emph{left} are given at the bottom of the table. }
\label{lines}
\centering
\begin{tabular}{lcccccccllllrr}
\cline{1-14}\\
 & \multicolumn{2}{c}{Inner Region} & \multicolumn{2}{c}{Outer Region} & \multicolumn{2}{c}{NW Knot} & & & & & & & \\
$\lambda_{\rm obs}$ & $F(\lambda)$ & $I(\lambda)$ & $F(\lambda)$ & $I(\lambda)$ & $F(\lambda)$ & $I(\lambda)$ & Ion & Mult. & $\lambda_{\rm lab}$ & Lower & Upper & $g_{1}$ & $g_{2}$\\
({\AA}) & & & & & & & & & ({\AA}) & & & & \\
\cline{1-14}\\
3726.2$^{a}$ &   47.8  &   75.2$\pm$2.5   &  14.7  &   23.1$\pm$1.5   &  347 & 545$\pm$34  &  [O~{\sc ii}] &       & 3727 &   2p$^{3}$ $^{4}$S$^{\rm o}$ & 2p$^{3}$ $^{2}$D$^{\rm o}$ &  4 & 4\\
3749.5 &    2.11 &   3.30$\pm$0.35  &        &                  &          &               &  H~{\sc i}    &  H12  & 3750 &   2p $^{2}$P$^{\rm o}$       & 12d $^{2}$D                &  8 &  \\
3757.8$^{b}$ &    1.87 &   2.91$\pm$0.32  &   4.65 &   7.2$\pm$0.8  &      &             &  O~{\sc iii}  &  V2   & 3757 &   3s $^{3}$P$^{\rm o}$       &  3p $^{3}$D                &  1 & 3\\
3769.9 &    2.89 &   4.5$\pm$0.5  &        &                  &          &               &  H~{\sc i}    &  H11  & 3770 &   2p $^{2}$P$^{\rm o}$       & 11d $^{2}$D                &  8 &  \\
3797.1 &    3.78 &   5.8$\pm$0.6  &        &                  &          &               &  H~{\sc i}    &  H10  & 3798 &   2p $^{2}$P$^{\rm o}$       & 10d $^{2}$D                &  8 &  \\
3815.9 &    1.08 &   1.7$\pm$0.4  &  12.9  &   20$\pm$5   &          &               &  He~{\sc ii}  &       & 3813 &   4f $^{2}$F$^{\rm o}$       & 19g $^{2}$G                & 32 &  \\
3834.4 &    5.19 &   7.9$\pm$0.6  &        &                  &          &               &  H~{\sc i}    &  H9   & 3835 &   2p $^{2}$P$^{\rm o}$       &  9d $^{2}$D                &  8 &  \\
3868.0 &   50.8  &   76.5$\pm$1.3   &  29.5  &   44.4$\pm$1.8   &          &               & [Ne~{\sc iii}]&       & 3868 &   2p$^{4}$ $^{3}$P           & 2p$^{4}$ $^{1}$D           &  5 & 5\\
3888.0$^{c}$ &   12.3  &   18.5$\pm$0.7   &  12.6  &   18.9$\pm$0.7   &    &               &  H~{\sc i}    &  H8   & 3889 &   2p $^{2}$P$^{\rm o}$       &  8d $^{2}$D                &  8 &  \\
3922.0 &    0.52 &   0.78$\pm$0.21  &        &                  &          &               &  He~{\sc ii}  &       & 3923 &   4f $^{2}$F$^{\rm o}$       & 15g $^{2}$G                & 32 &  \\
3967.5$^{d}$ &   26.5  &   38.6$\pm$1.2   &  18.7  &   27.3$\pm$1.8   &   80.6 & 117$\pm$8 & [Ne~{\sc iii}]&       & 3967 &   2p$^{4}$ $^{3}$P           & 2p$^{4}$ $^{1}$D           &  3 & 5\\
4024.8 &    1.68 &   2.40$\pm$0.35  &        &                  &          &               &  He~{\sc i}   &       & 4024 &   2p $^{1}$P$^{\rm o}$       &  7s $^{1}$S                &  1 & 3\\
4067.9 &    3.67 &   5.1$\pm$0.5  &        &                  &          &               &  [S~{\sc ii}] &       & 4068 &   3p$^{3}$ $^{4}$S$^{\rm o}$ & 2p$^{3}$ $^{2}$P$^{\rm o}$ &  4 & 4\\
4100.4$^{e}$ &   18.0  &   24.9$\pm$0.46  &  17.6  &   24.3$\pm$1.0  & 38.0 & 52.5$\pm$2.2 &  H~{\sc i}    &  H6   & 4101 &   2p $^{2}$P$^{\rm o}$       &  6d $^{2}$D                &  8 &72\\
4143.9 &   0.211 &   0.29$\pm$0.13  &        &                  &          &               &  He~{\sc i}   &  V53  & 4144 &   2p $^{1}$P$^{\rm o}$       &  6d $^{1}$D                &  3 & 5\\
4198.2 &    1.26 &   1.67$\pm$0.25  &        &                  &          &               &  He~{\sc ii}  &       & 4200 &   4f $^{2}$F$^{\rm o}$       & 11g $^{2}$G                & 32 &  \\
4226.7 &    0.24 &   0.32$\pm$0.10  &        &                  &          &               &  [Fe~{\sc v}] &       & 4227 &   3d$^{4}$ $^{5}$D           & 3d$^{4}$ $^{3}$H           &  9 & 9\\
4340.0 &   38.1  &   47.5$\pm$1.2   &  30.3  &   37.7$\pm$1.9   &   34.2   &  42.6$\pm$2.3 &  H~{\sc i}    &  H5   & 4340 &   2p $^{2}$P$^{\rm o}$       & 5d $^{2}$D                 &  8 &50\\
4362.5 &   14.6  &   18.0$\pm$0.6   &   7.82 &   9.6$\pm$0.6  &          &               &  [O~{\sc iii}]&       & 4363 &   2p$^{2}$ $^{1}$D           & 2p$^{2}$ $^{1}$S           &  5 & 1\\
4386.6 &    0.36 &   0.44$\pm$0.10  &        &                  &          &               &  He~{\sc i}   &  V15  & 4388 &   2p $^{1}$P$^{\rm o}$       & 5d $^{1}$D                 &  3 & 5\\
4470.9 &    2.31 &   2.72$\pm$0.21  &        &                  &          &               &  He~{\sc i}   &  V14  & 4471 &   2p $^{3}$P$^{\rm o}$       & 4d $^{3}$D                 &  9 &15\\
4541.2 &    2.94 &   3.35$\pm$0.19  &        &                  &          &               &  He~{\sc ii}  &       & 4541 &   4f $^{2}$F$^{\rm o}$       & 9g $^{2}$G                 & 32 &  \\
4640.8 &    1.92 &   2.1$\pm$0.8  &        &                  &          &               &  N~{\sc iii}  &  V2   & 4641 &   3p $^{2}$P$^{\rm o}$       & 3d $^{2}$D                 &  4 & 6\\
4685.6 &   88.6  &   95.0$\pm$2.1   & 117    &    126$\pm$6     &  132     &  142$\pm$6    &  He~{\sc ii}  &       & 4686 &   3d $^{2}$D                 & 4f $^{2}$F$^{\rm o}$       & 18 & 3\\
4711.3 &    8.25 &   8.8$\pm$0.8  &   7.76 &   8.2$\pm$1.1  &          &               &  [Ar~{\sc iv}]&       & 4711 &   3p$^{3}$ $^{4}$S$^{\rm o}$ & 3p$^{3}$ $^{2}$D$^{\rm o}$ &  4 & 6\\
4725.6 &    1.84 &   1.9$\pm$0.7  &        &                  &          &               &  [Ne~{\sc iv}]&       & 4724 &   2p$^{3}$ $^{2}$D$^{\rm o}$ & 2p$^{3}$ $^{2}$P$^{\rm o}$ &  4 & 6\\
4740.2 &    5.80 &   6.09$\pm$0.25  &   5.83 &   6.1$\pm$0.5  &          &               &  [Ar~{\sc iv}]&       & 4740 &   3p$^{3}$ $^{4}$S$^{\rm o}$ & 3p$^{3}$ $^{2}$D$^{\rm o}$ &  4 & 4\\
4861.6$^{f}$ &  100    &    100           & 100    &    100           &  100     &  100    &  H~{\sc i}    &  H4   & 4861 &   2p $^{2}$P$^{\rm o}$       & 4d $^{2}$D                 &  8 &32\\
4921.8 &    0.70 &   0.7$\pm$0.5  &        &                  &          &               &  He~{\sc i}   &  V48  & 4922 &   2p $^{1}$P$^{\rm o}$       & 4d $^{1}$D                 &  3 & 5\\
4959.3 &  252    &    243$\pm$3     & 168    &    162$\pm$2     &  408.31  &  394$\pm$8    &  [O~{\sc iii}]&       & 4959 &   2p$^{2}$ $^{3}$P           & 2p$^{2}$ $^{1}$D           &  3 & 5\\
5007.0 &  775    &    735$\pm$5     & 516    &    489$\pm$4     & 1109.88  & 1053$\pm$11   &  [O~{\sc iii}]&       & 5007 &   2p$^{2}$ $^{3}$P           & 2p$^{2}$ $^{1}$D           &  5 & 5\\
5145.5 &   0.30  &   0.27\,:     &        &                  &          &               &  [Fe~{\sc vi}]&       & 5145 &   3d$^{3}$ $^{4}$F           & 3d$^{3}$ $^{2}$G           &  8 & 8\\
5176.9 &   0.25  &   0.22\,:     &        &                  &          &               &  [Fe~{\sc vi}]&       & 5176 &   3d$^{3}$ $^{4}$F           & 3d$^{3}$ $^{2}$G           & 10 &10\\
5198.7$^{g}$ &    2.02 &   1.80$\pm$0.25  &        &                  &          &         &  [N~{\sc i}]  &       & 5198 &   2p$^{3}$ $^{4}$S$^{\rm o}$ & 2p$^{3}$ $^{2}$D$^{\rm o}$ &  4 & 4\\
5275.9 &   0.21  &   0.18\,:     &        &                  &          &               &  [Fe~{\sc vi}]&       & 5278 &   3d$^{3}$ $^{4}$F           & 3d$^{3}$ $^{4}$P           &  4 & 4\\
5335.6 &   0.30  &   0.26\,:     &        &                  &          &               &  [Fe~{\sc vi}]&       & 5335 &   3d$^{3}$ $^{4}$F           & 3d$^{3}$ $^{4}$P           &  4 & 2\\
5412.0 &   8.71  &   7.3$\pm$0.5  &   8.09 &   6.8$\pm$0.9  &          &               &  He~{\sc ii}  &       & 5411 &   4f $^{2}$F$^{\rm o}$       & 7g $^{2}$G                 & 32 &98\\
5483.8 &   0.14  &   0.11\,:     &        &                  &          &               &  [Fe~{\sc vi}]&       & 5485 &   3d$^{3}$ $^{4}$F           & 3d$^{3}$ $^{4}$P           &  6 & 2\\
5516.4 &   0.56  &   0.46$\pm$0.14  &        &                  &          &               & [Cl~{\sc iii}]&       & 5517 &   3p$^{3}$ $^{4}$S$^{\rm o}$ & 3p$^{3}$ $^{2}$D$^{\rm o}$ &  4 & 6\\
5538.3 &   0.74  &   0.60$\pm$0.14  &        &                  &          &               & [Cl~{\sc iii}]&       & 5537 &   3p$^{3}$ $^{4}$S$^{\rm o}$ & 3p$^{3}$ $^{2}$D$^{\rm o}$ &  4 & 4\\
5630.0 &   0.14  &   0.11\,:     &        &                  &          &               &  [Fe~{\sc vi}]&       & 5631 &   3d$^{3}$ $^{4}$F           & 3d$^{3}$ $^{4}$P           &  8 & 4\\
5678.1 &   0.25  &   0.20\,:     &        &                  &          &               &  [Fe~{\sc vi}]&       & 5677 &   3d$^{3}$ $^{4}$F           & 3d$^{3}$ $^{4}$P           & 10 & 6\\
5721.4 &   0.90  &   0.70$\pm$0.17  &        &                  &          &               & [Fe~{\sc vii}]&       & 5721 &   3d$^{2}$ $^{3}$F           & 3d$^{2}$ $^{1}$D           &  5 & 5\\
5755.4 &    5.73 &   4.42$\pm$0.19  &        &                  &          &               &  [N~{\sc ii}] &       & 5755 &   2p$^{2}$ $^{1}$D           & 2p$^{2}$ $^{1}$S           &  5 & 1\\
5876.2 &  12.3   &   9.2$\pm$0.5  &        &                  &          &               &  He~{\sc i}   &       & 5876 &   2p $^{3}$P$^{\rm o}$       & 3d $^{3}$D                 &  9 &15\\
6037.9 &   0.18  &   0.13$\pm$0.14  &        &                  &          &               &  He~{\sc ii}  &       & 6036 &   5g $^{2}$G                 & 21h $^{2}$H$^{\rm o}$      & 50 &  \\
6072.8 &   0.24  &   0.18$\pm$0.16  &        &                  &          &               &  He~{\sc ii}  &       & 6074 &   5g $^{2}$G                 & 20h $^{2}$H$^{\rm o}$      & 50 &  \\
6086.2 &   1.61  &   1.16$\pm$0.15  &        &                  &          &               & [Fe~{\sc vii}]&       & 6086 &   3d$^{2}$ $^{3}$F           & 3d$^{2}$ $^{1}$D           &  7 & 5\\
6101.3 &   0.42  &   0.30$\pm$0.10  &        &                  &          &               &  [K~{\sc iv}] &       & 6102 &   3p$^{4}$ $^{3}$P           & 3p$^{4}$ $^{1}$D           &  5 & 5\\
6117.9 &   0.32  &   0.23$\pm$0.13  &        &                  &          &               &  He~{\sc ii}  &       & 6118 &   5g $^{2}$G                 & 19h $^{2}$H$^{\rm o}$      & 50 &  \\
6169.9 &   0.31  &   0.22$\pm$0.12  &        &                  &          &               &  He~{\sc ii}  &       & 6170 &   5g $^{2}$G                 & 18h $^{2}$H$^{\rm o}$      & 50 &  \\
6232.5 &   0.49  &   0.34$\pm$0.14  &        &                  &          &               &  He~{\sc ii}  &       & 6234 &   5g $^{2}$G                 & 17h $^{2}$H$^{\rm o}$      & 50 &  \\
6300.3 &   8.52  &   5.89$\pm$0.25  &        &                  &          &               &  [O~{\sc i}]  &       & 6300 &   2p$^{4}$ $^{3}$P           & 2p$^{4}$ $^{1}$D           &  5 & 5\\
6311.8$^{h}$ &   5.02  &   3.47$\pm$0.14  &   4.51 &   3.11$\pm$0.19  &          &         &  [S~{\sc iii}]&       & 6312 &   3p$^{2}$ $^{1}$D           & 3p$^{2}$ $^{1}$S           &  5 & 1\\
6363.6 &   2.68  &   1.83$\pm$0.11  &        &                  &          &               &  [O~{\sc i}]  &       & 6363 &   2p$^{4}$ $^{3}$P           & 2p$^{4}$ $^{1}$D           &  5 & 3\\
\\
\cline{1-14}
\end{tabular}
\end{table*}

\addtocounter{table}{-1}
\begin{table*}
\caption{Continued.}
\label{lines}
\centering
\begin{tabular}{lcccccccllllrr}
\cline{1-14}\\
 & \multicolumn{2}{c}{Inner Region} & \multicolumn{2}{c}{Outer Region} & \multicolumn{2}{c}{NW Knot} & & & & & & & \\
$\lambda_{\rm obs}$ & $F(\lambda)$ & $I(\lambda)$ & $F(\lambda)$ & $I(\lambda)$ & $F(\lambda)$ & $I(\lambda)$ & Ion & Mult. & $\lambda_{\rm lab}$ & Lower & Upper & $g_{1}$ & $g_{2}$\\
({\AA}) & & & & & & & & & ({\AA}) & & & & \\
\cline{1-14}\\
6405.5 &  0.72  &   0.49$\pm$0.11  &        &                  &        &               & He~{\sc ii}  &       & 6406 & 5g $^{2}$G           & 15h $^{2}$H$^{\rm o}$ & 50 &  \\
6434.1 &  2.70  &   1.82$\pm$0.12  &   2.58 &   1.74$\pm$0.17  &        &               & [Ar~{\sc v}] &       & 6435 & 3p$^{2}$ $^{3}$P     & 3p$^{2}$ $^{1}$D      &  3 & 5\\
6526.1 &  0.75  &   0.50$\pm$0.12  &        &                  &        &               & [N~{\sc ii}] &       & 6527 & 2p$^{2}$ $^{3}$P     & 2p$^{2}$ $^{1}$D      &  1 & 5\\
6547.5 &  77.1  &     51$\pm$5     &  23.1  &   15.2$\pm$2.3   &  219   &   144$\pm$26  & [N~{\sc ii}] &       & 6548 & 2p$^{2}$ $^{3}$P     & 2p$^{2}$ $^{1}$D      &  3 & 5\\
6561.9 & 468    &    339$\pm$20    & 387    &    255$\pm$19    &  585   &   385$\pm$32  & H~{\sc i}    &       & 6563 & 2p $^{2}$P$^{\rm o}$ & 3d $^{2}$D            &  8 &18\\
6582.9 & 248    &    162$\pm$5     &  69.9  &   45.8$\pm$2.3   &  540   &   354$\pm$22  & [N~{\sc ii}] &       & 6583 & 2p$^{2}$ $^{3}$P     & 2p$^{2}$ $^{1}$D      &  5 & 5\\
6678.3$^{i}$ &   4.13 &    2.7$\pm$0.4   &        &                  &        &         & He~{\sc i}   &   V46 & 6678 & 2p $^{1}$P$^{\rm o}$ & 3d $^{1}$D            &  3 & 5\\
6715.7 &   7.51 &   4.81$\pm$0.22  &   3.58 &   2.29$\pm$0.21  &        &               & [S~{\sc ii}] &       & 6716 & 2p$^{3}$ $^{4}$S$^{\rm o}$ & 2p$^{3}$ $^{2}$D$^{\rm o}$ &  4 & 6\\
6730.5 &  12.7  &   8.11$\pm$0.25  &   6.27 &   4.00$\pm$0.25  &   47.6 &  30.4$\pm$2.4 & [S~{\sc ii}] &       & 6731 & 2p$^{3}$ $^{4}$S$^{\rm o}$ & 2p$^{3}$ $^{2}$D$^{\rm o}$ &  4 & 4\\
6820.9 &  0.52  &   0.33$\pm$0.12  &        &                  &        &               & [Fe~{\sc v}] &       & 6819 & 3d$^{4}$ $^{3}$P4    & 3d$^{4}$ $^{1}$S4     &  3 & 1\\
6889.8 &   1.13 &   0.70$\pm$0.14  &        &                  &        &               & He~{\sc ii}  &       & 6891 & 5g $^{2}$G           & 12h $^{2}$H$^{\rm o}$ & 50 &  \\
7004.8 &   6.43 &   3.90$\pm$0.20  &   3.97 &   2.41$\pm$0.24  &        &               & [Ar~{\sc v}] &       & 7006 & 3p$^{2}$ $^{3}$P     & 3p$^{2}$ $^{1}$D      &  5 & 5\\
7065.2 &   7.22 &   4.33$\pm$0.24  &   4.00 &   2.40$\pm$0.26  &        &               & He~{\sc i}   &       & 7065 & 2p $^{3}$P$^{\rm o}$ & 3s $^{3}$S            &  9 & 3\\
\\
$\log\,F$(H$\beta$)$^{j}$ & $-$11.85 & & $-$13.61 & & $-$14.47 & & & & & & & & \\
\\
\cline{1-14}
\end{tabular}
\begin{description}
\item[$^{a}$] A blend of the [O~{\sc ii}] $\lambda\lambda$3726,\,3729 lines.

\item[$^{b}$] Blended with the O~{\sc iii} 3760 (3s\,$^{3}$P$^{\rm o}_{2}$
-- 3p\,$^{3}$D$_{3}$) line.

\item[$^{c}$] Blended with the He~{\sc i} $\lambda$3888 (2s\,$^{3}$S --
3p\,$^{3}$P$^{\rm o}$) line.

\item[$^{d}$] Blended with the H~{\sc i} $\lambda$3970 (2p\,$^{2}$P$^{\rm
o}$ -- 7d\,$^{2}$D) line.

\item[$^{e}$] Blended with the N~{\sc iii} $\lambda$4103 (3s\,$^{2}$S$_{1/2}$
-- 3p\,$^{2}$P$^{\rm o}_{1/2}$) line.

\item[$^{f}$] Blended with He~{\sc ii} $\lambda$4859 line, whose flux
contribution is negligible.  The same happens to H$\alpha$, whose is
blended with He~{\sc ii} $\lambda$6560.  See the text for details.

\item[$^{g}$] Blended with the [N~{\sc i}] $\lambda$5200
(2p$^{3}$\,$^{4}$S$^{\rm o}_{3/2}$ -- 2p$^{3}$\,$^{2}$D$^{\rm o}_{5/2}$)
line.

\item[$^{h}$] Corrected for the flux of the blended He~{\sc ii}
$\lambda$6311 (5g\,$^{2}$G -- 16h\,$^{2}$H$^{\rm o}$) line.

\item[$^{i}$] Corrected for the flux of the blended He~{\sc ii}
$\lambda$6683 (5g\,$^{2}$G -- 13h\,$^{2}$H$^{\rm o}$) line.

\item[$^{j}$] erg\,cm$^{-2}$\,s$^{-1}$ in our extracted spectra.
\end{description}
\end{table*}

\section{Spectral analysis}

\begin{figure*}
\begin{center}
\epsfig{file=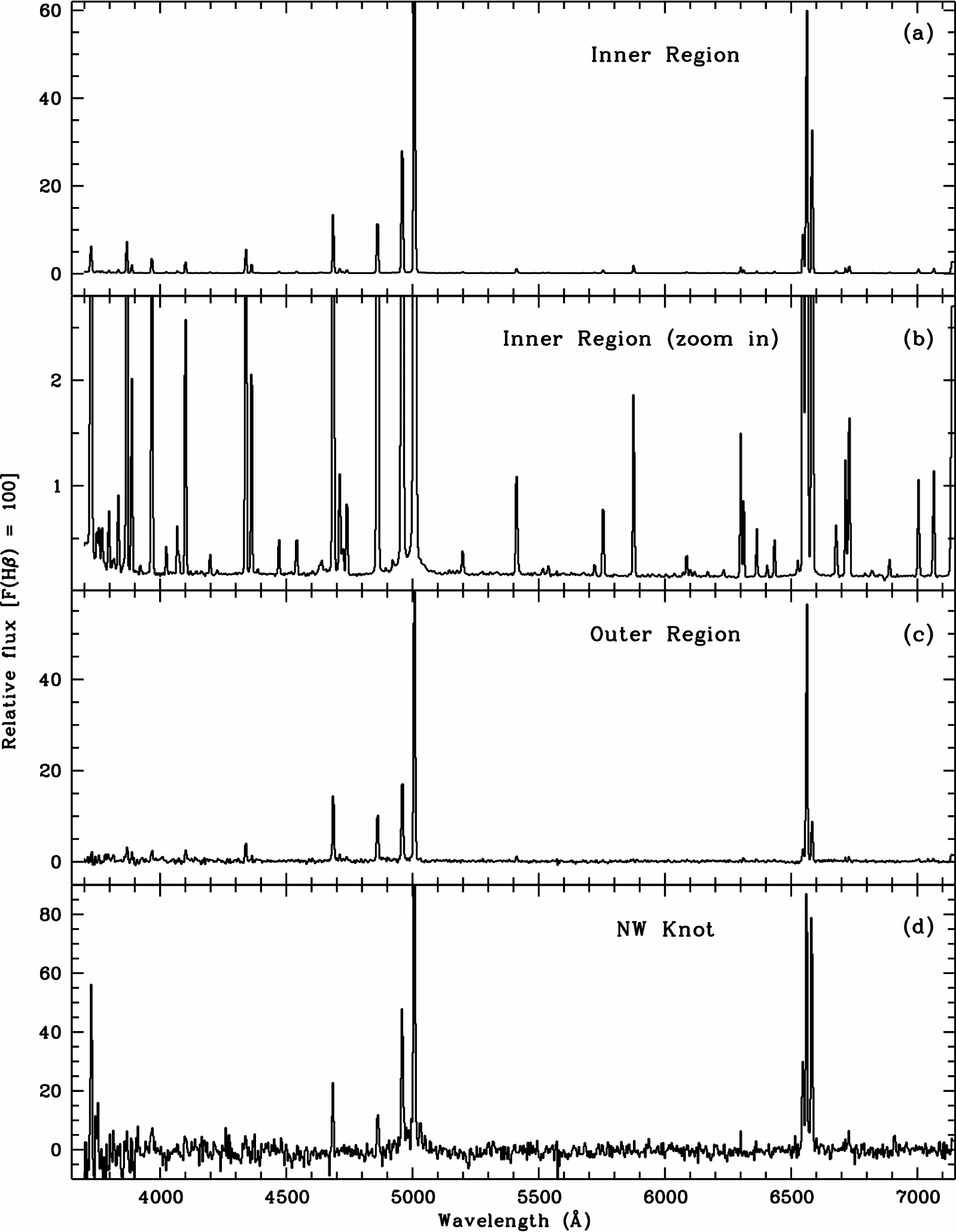,width=16.75cm,angle=0}
\caption{
One-dimensional intermediate-dispersion optical spectra of the inner region
(a and b), outer region (c), and NW knot (d) of Hu\,1-2 in the spectral
range 3700--7100~\AA.  Panel (b) is a zoom-in view of panel (a)
highlighting the weakest emission lines detected in the inner-region
spectrum.  All spectra have been normalized such that H$\beta$ has an
integrated flux of 100.  Extinction has not been corrected for.  Spectra
are scaled to fit the peak intensity of the H$\alpha$ line.  Differences
in the relative strengths of the [O~{\sc ii}] and [N~{\sc ii}] lines in the
three nebular regions are obvious. } 
\label{spec}
\end{center}
\end{figure*}

One-dimensional (1-D) spectra of the bright inner region, the bipolar 
lobes, and the bipolar knots were extracted with the spatial extents 
shown in the H$\alpha$ image in Figure~\ref{grey_img}-\emph{left}.  The 
contamination by a background star with absorptions in the hydrogen 
Balmer lines renders of little use of the spectrum of the SE knot.  The 
1-D spectra showed scattered light from the bright inner region of the 
H$\beta$+[O~{\sc iii}] and H$\alpha$+[N~{\sc ii}] lines.  As a result, 
the 1-D spectra displayed a pedestal of emission under the [O~{\sc iii}] 
and H$\alpha$+[N~{\sc ii}] lines.  We removed this dispersed light by 
fitting a long-pass Gaussian profile to the nebular continuum around 
these lines.  After this correction, the [O~{\sc iii}] 
$\lambda$5007/$\lambda$4959 and [N~{\sc ii}] $\lambda$6583/$\lambda$6548 
line ratios measured from the spectra are both close to their theoretical 
value of $\sim$3.  The corrected 1-D spectra are shown in Figure~\ref{spec}. 
We also integrated total fluxes of the [O~{\sc iii}] $\lambda$5007, [N~{\sc 
ii}] $\lambda$6583, and H$\alpha$ emission lines for the main nebula and 
estimated the surface brightnesses.  The brightness values were used to 
calibrate the ALFOSC narrow-band images of Hu\,1-2 
(Section~\ref{observe:1}).

\subsection{Physical conditions}

The 1-D intermediate-dispersion, normalized spectra of the inner region, 
outer region, and NW knot of Hu\,1-2 are shown in Figure~\ref{spec}.  
Although at intermediate-dispersion, the long slit spectrum of Hu\,1-2 
presented here proves to be very deep.  The signal-to-noise ratio (S/N) 
of the spectrum of the inner region is particularly high, and thus many 
weak emission lines with fluxes lower than 1\% of H$\beta$ are detected 
(see Figure~\ref{spec}).  The emission lines detected in the spectra of 
the three regions are compiled in Table~\ref{lines}. 

We derived a logarithmic extinction parameter, $c$(H$\beta$), of $0.61$ in 
the inner region of Hu\,1-2 using the observed H$\alpha$/H$\beta$ line
ratio. 
Here the theoretical H$\alpha$/H$\beta$ line ratio was adopted
from Storey \& Hummer \cite{sh95}, assuming an electron temperature of
10\,000~K and a density of 10$^{4}$~cm$^{-3}$. 
Our $c$(H$\beta$) value agrees with 0.60 given by Pottasch et al.\ 
(\citealt{pot03}; also Hyung, Pottasch \& Feibelman \citealt{hyu04}), who 
carried out echelle spectroscopy at the center of Hu\,1-2 with a slit 
entrance 1\farcs2$\times$4\arcsec, i.e., also the inner region of Hu\,1-2. 
Earlier observations by Peimbert \& Torres-Peimbert \cite{ptp87} and Aller 
\& Czyzak \cite{ac79} gave an extinction value of 0.64 and 0.61, 
respectively. 
The $c$(H$\beta$) values derived for the outer region and the 
NW knot are $\sim$0.40 and 0.90, respectively.  Given the more accurate 
emission-line measurements for the inner region, we adopted 0.61 as the 
extinction in all three regions of Hu\,1-2, 
although we cannot totally rule out the possibility that extinction 
in the NW knot could be higher. 
The extinction law of Whitford \cite{whit58} was used to apply this 
correction.  
We found that the use of different extinction laws (e.g., Whitford 
\citealt{whit58}; Savage \& Mathis \citealt{sm79}; Cardelli, Clayton 
\& Mathis \citealt{ccm89}) results in differences in the 
extinction-corrected fluxes of emission lines in the optical range 
from $\sim$3700\,{\AA} to 7100\,{\AA} lower than 5\%.

The extinction-corrected relative line intensities, $I$($\lambda$), 
together with the estimated errors, are given in Table~\ref{lines}. 
Errors in the line intensities were estimated from multiple measurements 
of the observed fluxes, through the almost linear relation between the 
two quantities, 
\begin{equation}
\label{eq2}
I(\lambda) = 10^{c({\rm H}\beta)\,f(\lambda)} F(\lambda), 
\end{equation}
where $f$($\lambda$) is the reddening function, normalized to 
$f$(H$\beta$) = 0, adopted from Whitford \cite{whit58}.  Uncertainties 
in the extinction function are supposed to be negligible, as discussed
in the previous paragraph.  Flux calibration, which can affect the 
observed fluxes and consequently the extinction-corrected relative line 
intensities, was supposed to be accurate.

The [O~{\sc iii}] $\lambda$5007 line was saturated in the long-exposure 
(1800\,s) spectrum of the inner region, and the flux from the short-exposure 
(180\,s) spectrum was thus used in Table~\ref{lines}.  The He~{\sc ii}
$\lambda$6560 (4f\,$^{2}$F$^{\rm o}$ -- 6g\,$^{2}$G) line is blended with 
H$\alpha$.  Using the theoretical line ratios of the hydrogenic ions (Storey
\& Hummer \citealt{sh95}) and the He~{\sc ii} $\lambda$4686 line fluxes 
measured in our spectra, we estimated that the He~{\sc ii} $\lambda$6560 
line contributes $\sim$3\% and 6\% to the total flux of H$\alpha$ in 
the inner and outer region of Hu\,1-2, respectively.  As a consequence, the 
corrected fluxes of H$\alpha$ will result in an increase in $c$(H$\beta$) 
by $\sim$9\% and 14\% in the two regions. That will cause negligible 
changes in the extinction-corrected line intensities, given that the 
extinction in Hu\,1-2 is low and that the inner-region extinction was adopted
for the whole nebula.

The electron temperatures and densities derived for the inner and outer 
regions using the detected emission lines are presented in 
Table~\ref{temden}.  The [O~{\sc iii}] nebular-to-auroral line ratio yielded 
a temperature of 16\,800~K for the inner region, and 15\,200~K for the outer 
region.  The [N~{\sc ii}] line ratio yielded a temperature of $\sim$13\,000~K
in the inner region, but the temperature-sensitive $\lambda$5755 auroral line 
was not detected in the outer region.  We adopted the above two temperatures 
for the high- and low-excitation regions in Hu\,1-2.  An averaged electron 
density of 5700~cm$^{-3}$ in the inner region was derived from the [S~{\sc 
ii}] $\lambda$6716/$\lambda$6731 and [Cl~{\sc iii}] 
$\lambda$5517/$\lambda$5537 line ratios.  It agrees with the density
5770~cm$^{-3}$ in the outer region, as yielded by the [S~{\sc ii}] lines 
(Table~\ref{temden}).  The [Cl~{\sc iii}] lines were not detected in the 
outer region.

\subsection{Chemical abundances}

\subsubsection{Ionic abundances}

Using the electron temperatures defined for the high- and low-excitation
zones, we derived the ionic abundances of helium and heavy elements 
presented in Table~\ref{ionic}.  
The He~{\sc i} effective recombination coefficients used for abundance 
determination were adopted from Benjamin, Skillman \& Smits \cite{bss99}.
For the inner region, we adopted the He$^{+}$/H$^{+}$ abundance ratio 
derived from the $\lambda$5876 line which is the strongest He~{\sc i} 
line.  For the outer region, the He$^{+}$/H$^{+}$ abundance derived from 
the $\lambda$7065 line was adopted.  
No He~{\sc i} lines were detected in the spectrum of the NW knot.

The He$^{2+}$/H$^{+}$ abundance was derived from the He~{\sc ii} 
$\lambda$4686 line, which is very strong in all the three regions 
of Hu\,1-2 (Figure~\ref{spec}).  
The extinction-corrected flux of the He~{\sc ii} $\lambda$4686 line in the 
outer region is even higher than that of the H$\beta$ line by $\sim$25\%, 
in consistency with the low intensities of the [N~{\sc ii}] and [O~{\sc ii}] 
nebular lines.  
Note, however, that the relative strengths of the [O~{\sc ii}] $\lambda$3727 
(a blend of the $\lambda\lambda$3726,\,3729 doublet), [N~{\sc ii}] 
$\lambda\lambda$6548,\,6583, and He~{\sc ii} $\lambda$4686 lines in the NW 
knot are higher than those in the inner and outer regions 
(Table~\ref{lines}).
The He/H abundance is a sum of the 
He$^{+}$/H$^{+}$ and He$^{2+}$/H$^{+}$ ratios and are given in 
Table~\ref{elemental}.  The He/H ratio in the inner region (0.144) is slightly
lower than that in the outer region (0.159), but a little higher than the NW 
knot, where no He~{\sc i} line was detected.  The absence of He~{\sc i} lines
in the NW knot probably indicates that helium in the NW knot is all doubly 
ionized, although the relatively low S/N's of the spectrum in this region 
may hinder the detection of the He~{\sc i} lines.  
Given that measurements of the He~{\sc i} $\lambda$7065 line in the 
outer region of Hu\,1-2 have relatively large uncertainties, the He/H 
abundance in the outer region is considered to agree with that in the 
inner region within the errors.


The N$^{+}$/H$^{+}$ abundance derived from the [N~{\sc ii}] $\lambda$6583 
line was adopted for the three regions because measurements of the relatively
weaker $\lambda$6548 line were much affected by the nearby H$\alpha$ line. 
The Ne$^{2+}$/H$^{+}$ abundances derived from the [Ne~{\sc iii}] 
$\lambda$3869 line were adopted because the [Ne~{\sc iii}] $\lambda$3967 line
was blended with the H~{\sc i} $\lambda$3970 line.  
The [S~{\sc iii}] $\lambda$6312 line was observed in the inner and outer 
regions.  It was blended with the He~{\sc ii} $\lambda$6311 (5g\,$^{2}$G -- 
16h\,$^{2}$H$^{\rm o}$) line, whose flux contribution was estimated from the
observed $\lambda$4686 line using the hydrogenic atomic model of Storey \& 
Hummer \cite{sh95}.  The corrected flux of the [S~{\sc iii}] $\lambda$6312 
line was used to derive S$^{2+}$/H$^{+}$.  The Ar$^{3+}$/H$^{+}$ ratio 
derived from the [Ar~{\sc iv}] $\lambda$4740 line was adopted for Hu\,1-2, 
as the $\lambda$4711 line was blended with [Ne~{\sc iv}] lines.

The abundance errors following the ionic abundance ratios in 
Table~\ref{ionic} were mainly propagated from the measurement errors 
of the extinction-corrected relative line intensities given in 
Table~\ref{lines}.  Both the [O~{\sc iii}] $\lambda$4363 and the 
[N~{\sc ii}] $\lambda$5755 temperature-sensitive auroral lines were 
well detected in the inner region of Hu\,1-2, and the electron 
temperatures derived from these ions were adopted for all the three 
regions when calculating ionic abundance ratios (Table~\ref{ionic}). 
Thus the errors in the electron temperature were small 
(Table~\ref{temden}) and not considered in the ionic abundance 
calculations.

\subsubsection{Elemental abundances}

The elemental abundances of He, N, O, Ne, S, and Ar in 
Table~\ref{elemental} were calculated using the ionization 
correction factor (ICF) method of Kingsburgh \& Barlow \cite{kb94}.  
No elemental abundances were calculated for the NW knot because the 
He~{\sc i} line needed for ICF calculations was not detected in this 
region.  
The He/H abundance in the inner region of Hu\,1-2 derived in this paper 
is $\sim$10\% higher than that derived by Pottasch et al.\ \cite{pot03} 
and Hyung, Pottasch \& Feibelman \cite{hyu04}, but it agrees with that 
of Peimbert \& Torres-Peimbert \cite{ptp87} 
and Peimbert, Luridiana \& Torres-Peimbert \cite{pltp95} 
within the errors.  Meanwhile, the nitrogen abundance derived for the 
inner region by us is $\sim$50\% higher than that of the outer region, 
and both are lower than that of previous observations.  
The oxygen, neon and sulfur abundances in this paper generally agree 
with those of Pottasch et al.\ \cite{pot03} and Hyung, Pottasch \& 
Feibelman \cite{hyu04}, while our argon abundance is slightly lower. 
However, uncertainties in abundances were not indicated in the two 
previous studies.  We also notice that the inner-region abundances of 
N, O, and Ne are higher than those in the outer region.  

Errors in the brackets following the total elemental abundances 
in Table~\ref{elemental} (the first two columns) were estimated mainly 
based on the errors in the ionic abundances through a simple propagation 
paradigm. 
For helium, the error is contributed by uncertainties in 
the He$^{+}$/H$^{+}$ and He$^{2+}$/H$^{+}$ ratios.  For heavy elements, 
the errors can also be introduced by the use of ICFs.  This source of 
error is negligible for oxygen, whose ICF is always close to unity. 
For other heavy elements, uncertainties introduced by ICFs could be 
significant.  The nitrogen and neon abundances were derived based on 
the ionic and elemental abundances of oxygen, and thus are reliable. 
The total sulfur abundance is generally quite uncertain, as usually 
only the [S~{\sc ii}] lines are well observed for this element. Although 
the [S~{\sc iii}] $\lambda$6312 line was also detected in the spectrum 
of Hu\,1-2 (Table~\ref{lines}), it arises from an auroral transition 
(3p$^{2}$\,$^{1}$D$_{2}$ -- 3p$^{2}$\,$^{1}$S$_{0}$), hence it is 
particularly temperature sensitive.  Besides, the [S~{\sc iii}] 
$\lambda$6312 line is also blended with the He~{\sc ii} $\lambda$6311 
(5g\,$^{2}$G -- 16h\,$^{2}$H$^{\rm o}$) line.  Uncertainties and 
systematic errors in the ICFs are difficult to define, and thus they 
were not considered for the error estimate of our object.  The actual 
uncertainties in the elemental abundances of nitrogen, neon, sulfur and 
argon in Table~\ref{elemental} must be regarded as lower limits of the 
real abundance uncertainties.


Average abundances of Type~I, bulge, and disc PNe as well as the 
solar abundances are also presented in Table~\ref{elemental} for 
comparison.  
The He/H abundances of the inner and outer regions of Hu\,1-2 are 
both higher than 0.14, whereas the N/O ratios of the two regions 
are both close to 0.9.  
The He/H and N/O abundance ratios, generally in line with those of Pottasch 
et al.\ \cite{pot03} and Hyung, Pottasch \& Feibelman \cite{hyu04}, indicate 
that Hu\,1-2 belongs to Type~I PNe (Peimbert \& Torres-Peimbert 
\citealt{ptp83}).  On the other hand, the elemental abundances of oxygen, 
neon, sulfur, and argon in Hu\,1-2 are lower than the average abundances 
of bulge and disc PNe (see Table~\ref{elemental}).  This difference 
in abundances is real for the heavy elements, given the uncertainties 
presented in Table~\ref{elemental} (see the discussion above). 
Since these four $\alpha$ elements were not supposed to be produced 
or depleted in the central star during its evolution\footnote{
This is not the exact case for, at least, the production of neon 
in some low- and intermediate-mass stars, as analyzed by, e.g., Wang \& 
Liu \cite{wl08} and Milingo et al.\ \cite{mil10}.  Karakas \& Lattanzio  
\cite{kar03} also predicted that neon can be produced in a narrow 
range of stellar mass. }, their very low abundances reflect those 
of the interstellar medium from which the progenitor star of Hu\,1-2 
formed.  Indeed the Ne/H vs.\ O/H, S/H vs.\ O/H, and Ar/H vs.\ O/H 
abundance patterns of $\alpha$ elements in Hu\,1-2 generally agree 
with those established for H~{\sc ii} regions and blue compact galaxies 
by Milingo et al.\ (2010; Figures~1--10 therein). 
In these figures, the loci of Hu\,1-2 are found at the low 
abundance tail of the samples of Type~I PNe (Milingo et al.\ 2010) and 
Galactic anticenter PNe (Kwitter et al.\ 2010).



\begin{table}
\begin{minipage}{65mm}
\centering
\caption{Electron density and temperature of Hu\,1-2. }
\label{temden}
\begin{tabular}{lrr}
\hline
 & Inner region & Outer region\\
\hline
 & \multicolumn{2}{c}{$T_{\rm e}$ (K)}\\
$[$O~{\sc iii}$]$  & 16\,800$\pm$250 & 15\,200$\pm$400\\
$[$N~{\sc ii}$]$   & 13\,000$\pm$400 &                \\
\\
 & \multicolumn{2}{c}{$N_{\rm e}$ (cm$^{-3}$)}\\
$[$S~{\sc ii}$]$   & 5100$\pm$1300  &  5770$\pm$2200\\
$[$Cl~{\sc iii}$]$ & 6400$\pm$2800  &               \\
$[$Ar~{\sc iv}$]$  & 3900$\pm$1600  &  3200$\pm$2500 \\
\hline
\end{tabular}
\end{minipage}
\end{table}

\begin{table*}
\begin{minipage}{115mm}
\centering
\caption{Ionic abundances$^{a}$ derived for the inner region, the outer 
region, and the NW knot of Hu\,1-2.}
\label{ionic}
\begin{tabular}{lccccc}
\hline
Ion & $\lambda$ & $T_\mathrm{e}\,^b$ & \multicolumn{3}{c}{Abundance}    \\
    & ({\AA})   & (K)                & \multicolumn{3}{c}{(X$^+$/H$^+$)}\\
\hline
    &       &          & Inner                 & Outer                 & Knot                  \\
\hline
He$^{+}$  & 4471 & 10\,000 & 0.052$\pm$0.004       &                       &                       \\
He$^{+}$  & 5876 & 10\,000 & 0.060$\pm$0.003       &                       &                       \\
He$^{+}$  & 7065 & 10\,000 & 0.087$\pm$0.005       & 0.048$\pm$0.005       &                       \\
He$^{2+}$ & 4686 & 16\,800 & 0.084$\pm$0.002       & 0.111$\pm$0.005       & 0.126$\pm$0.007       \\
N$^{+}$   & 5755 & 13\,000 & 1.70($\pm$0.25)$\times$10$^{-5}$ &                                  &                                 \\
N$^{+}$   & 6548 & 13\,000 & 1.66($\pm$0.10)$\times$10$^{-5}$ & 4.96($\pm$0.33)$\times$10$^{-6}$ &  4.7($\pm$0.4)$\times$10$^{-5}$ \\
N$^{+}$   & 6583 & 13\,000 & 1.79($\pm$0.12)$\times$10$^{-5}$ &  5.1($\pm$0.4)$\times$10$^{-6}$  & 3.93($\pm$0.30)$\times$10$^{-5}$\\
O$^{+}$   & 3727 & 13\,000 & 1.92($\pm$0.25)$\times$10$^{-5}$ &  5.9($\pm$0.8)$\times$10$^{-6}$  & 1.40($\pm$0.12)$\times$10$^{-4}$\\
O$^{2+}$  & 4363 & 16\,800 &  5.9($\pm$0.5)$\times$10$^{-5}$  & 3.16($\pm$0.24)$\times$10$^{-5}$ &                                 \\
O$^{2+}$  & 4959 & 16\,800 & 5.67($\pm$0.24)$\times$10$^{-5}$ & 3.78($\pm$0.15)$\times$10$^{-5}$ &  9.2($\pm$0.4)$\times$10$^{-5}$ \\
O$^{2+}$  & 5007 & 16\,800 & 5.90($\pm$0.25)$\times$10$^{-5}$ & 3.95($\pm$0.18)$\times$10$^{-5}$ &  8.5($\pm$0.4)$\times$10$^{-5}$ \\
Ne$^{2+}$ & 3869 & 16\,800 & 1.40($\pm$0.07)$\times$10$^{-5}$ &  8.1($\pm$0.6)$\times$10$^{-6}$  &                                 \\
Ne$^{2+}$ & 3967 & 16\,800 & 2.34($\pm$0.14)$\times$10$^{-5}$ & 1.66($\pm$0.14)$\times$10$^{-5}$ &  7.1($\pm$0.5)$\times$10$^{-5}$ \\
S$^{+}$   & 4068 & 13\,000 & 4.26($\pm$0.35)$\times$10$^{-7}$ &                                  &                                 \\
S$^{+}$   & 6716 & 13\,000 &  3.4($\pm$0.4)$\times$10$^{-7}$  & 1.62($\pm$0.24)$\times$10$^{-7}$ &                                 \\
S$^{+}$   & 6731 & 13\,000 & 3.28($\pm$0.30)$\times$10$^{-7}$ & 1.62($\pm$0.18)$\times$10$^{-7}$ & 1.23($\pm$0.14)$\times$10$^{-6}$\\
S$^{2+}$  & 6312 & 13\,000 & 2.66($\pm$0.35)$\times$10$^{-6}$ & 2.26($\pm$0.26)$\times$10$^{-6}$ &                                 \\
Cl$^{2+}$ & 5517 & 13\,000 &  3.7($\pm$0.4)$\times$10$^{-8}$  &                                  &                                 \\
Cl$^{2+}$ & 5537 & 13\,000 & 3.77($\pm$0.34)$\times$10$^{-8}$ &                                  &                                 \\
Ar$^{3+}$ & 4711 & 16\,800 &  4.8($\pm$0.9)$\times$10$^{-7}$  &  4.6($\pm$0.8)$\times$10$^{-7}$  &                                 \\
Ar$^{3+}$ & 4740 & 16\,800 & 3.27($\pm$0.50)$\times$10$^{-7}$ &  3.3($\pm$0.6)$\times$10$^{-7}$  &                                 \\
Ar$^{4+}$ & 6435 & 16\,800 &  2.9($\pm$0.4)$\times$10$^{-7}$  &  2.8($\pm$0.5)$\times$10$^{-7}$  &                                 \\
Ar$^{4+}$ & 7006 & 16\,800 & 2.31($\pm$0.36)$\times$10$^{-7}$ & 1.43($\pm$0.26)$\times$10$^{-7}$ &                                 \\
K$^{3+}$  & 6102 & 16\,800 &  8.1($\pm$2.1)$\times$10$^{-9}$  &                                  &                                 \\
\hline
\end{tabular}
\begin{description}
\item[$^{a}$] Number ratios relative to hydrogen. 
\item[$^{b}$] The effective recombination coefficients used for the
He$^{+}$/H$^{+}$ abundance calculations were adopted from Benjamin, 
Skillman \& Smits \cite{bss99}, where only three temperature cases, 
5000~K, 10\,000~K, and 20\,000~K are presented.  We have used the 
He~{\sc i} atomic data at 10\,000~K, which is close to the low-excitation 
temperature (13\,000~K) derived for Hu\,1-2.
\end{description}
\end{minipage}
\end{table*}

\begin{table*}
\caption{Elemental abundances derived for the inner and outer regions of 
Hu\,1-2.  Abundances from the literature are also presented for comparison.
}
\label{elemental}
\centering
\begin{tabular}{lllllllll}
\hline
Ele. & \multicolumn{8}{c}{Abundance (X/H)}\\
     & Inner$^{a}$ & Outer$^{b}$ & Pottasch$^{c}$ & Hyung$^{d}$ & PTP87$^{e}$ & Bulge$^{f}$ & Disk$^{g}$ & Solar$^{h}$\\
\hline
He & 0.144$\pm$0.013                 & 0.159$\pm$0.022                  & 0.127                & 0.130                & 0.147                & 0.105                 & 0.115                 & 0.085                \\
N  & 1.31($\pm$0.22)$\times$10$^{-4}$ & 8.7($\pm$1.7)$\times$10$^{-5}$   & 1.9$\times$10$^{-4}$ & 1.7$\times$10$^{-4}$ & 2.2$\times$10$^{-4}$ & 1.48$\times$10$^{-4}$ & 2.19$\times$10$^{-4}$ & 6.76$\times$10$^{-5}$\\
O  & 1.40($\pm$0.20)$\times$10$^{-4}$ & 1.01($\pm$0.14)$\times$10$^{-4}$ & 1.6$\times$10$^{-4}$ & 1.3$\times$10$^{-4}$ & 1.6$\times$10$^{-4}$ & 3.98$\times$10$^{-4}$ & 5.01$\times$10$^{-4}$ & 4.90$\times$10$^{-4}$\\
Ne & 3.3($\pm$0.7)$\times$10$^{-5}$   & 2.1($\pm$0.5)$\times$10$^{-5}$   & 4.9$\times$10$^{-5}$ & 3.0$\times$10$^{-5}$ & 3.6$\times$10$^{-5}$ & 9.77$\times$10$^{-5}$ & 1.35$\times$10$^{-4}$ & 8.51$\times$10$^{-5}$\\
S  & 4.2($\pm$1.3)$\times$10$^{-6}$   & 4.4($\pm$1.5)$\times$10$^{-6}$   & 4.2$\times$10$^{-6}$ & 3.5$\times$10$^{-6}$ &                      & 6.92$\times$10$^{-6}$ & 1.12$\times$10$^{-5}$ & 1.32$\times$10$^{-5}$\\
Ar & 7.8($\pm$2.2)$\times$10$^{-7}$   & 8.4($\pm$2.9)$\times$10$^{-7}$   & 1.1$\times$10$^{-6}$ & 1.1$\times$10$^{-6}$ & 7.9$\times$10$^{-7}$ & 1.58$\times$10$^{-6}$ & 2.19$\times$10$^{-6}$ & 2.51$\times$10$^{-6}$\\
\hline
\end{tabular}
\begin{description}
\item[$^{a}$] Elemental abundances of the inner region of Hu\,1-2.
\item[$^{b}$] Elemental abundances of the outer region of Hu\,1-2.
\item[$^{c}$] Pottasch et al. \cite{pot03}.
\item[$^{d}$] Hyung, Pottasch \& Feibelman \cite{hyu04}.
\item[$^{e}$] Peimbert \& Torres-Peimbert \cite{ptp87}.
\item[$^{f}$] Average abundances for 23 Galactic bulge PNe. Sample observed
by Wang \& Liu \cite{wl07} plus bulge PNe M\,1-42 and M\,2-36 analyzed by Liu
et al. \cite{liu01}.
\item[$^{g}$] Average abundances given by Wang \& Liu \cite{wl07} for 58
Galactic disc PNe. 
\item[$^{h}$] Solar values from Asplund et al. \cite{asp09}.
\end{description}
\end{table*}

\subsection{Could Hu\,1-2 be a halo PN?}

The abundances of the heavy elements of Hu\,1-2 are generally consistent 
with those of the halo PNe, given their comparatively large abundance 
scatter as opposed to the disc PNe (Howard, Henry \& McCartney 
\citealt{hhm97}).  Note, however, that the helium abundances of Hu\,1-2 
are slightly higher than those of the sample of halo PNe studied by 
Howard, Henry \& McCartney \cite{hhm97}.  It is thus pertinent to 
question whether Hu\,1-2 is a halo PN or not.

Halo PNe are characterized by their height above the Galactic plane, 
their kinematic characteristics, and/or their low metallicity relative 
to disc PNe.  Specifically Peimbert \cite{pei90} proposed $|z|\,>$ 
0.8~kpc, $v_{\rm pec}\,>$ 60 km\,s$^{-1}$, and $\log$(O/H)+12 $<$ 8.1 
as the criteria for halo PNe, where $z$ is the distance to the Galactic 
plane and $v_{\rm pec}$ is the peculiar radial velocity, which is the 
difference between the observed systemic radial velocity of a PN and 
its circular radial velocity ($v_{\rm pec}$ = $v_{\rm sys}$ $-$ $v_{\rm 
circ}$; Pe\~{n}a, Rechy-Garc\'{i}a \& Garc\'{i}a-Rojas \citealt{pena13}). 
For Hu\,1-2, at a Galactic latitude of $-$8\fdg8, its distance of 
3.5$^{+1.5}_{-0.8}$~kpc (Miranda et al.\ \citealt{mir12a}) implies a 
height of 0.55$^{+0.23}_{-0.13}$~kpc over the Galactic plane.  Its oxygen 
abundance is $\log$(O/H)+12 $\sim$8.1 (see Table~5).  Therefore, the 
oxygen abundance and its distance to the Galactic plane marginally 
support a halo nature.  
However, following the formulation by Pe\~{n}a, Rechy-Garc\'{i}a \& 
Garc\'{i}a-Rojas \cite[][Equation~1 therein]{pena13}, assuming a 
galactocentric distance of 8.3~kpc for the Sun, adopting a distance of 
3.5$^{+1.5}_{-0.8}$~kpc to Hu\,1-2, and taking into account the systemic 
velocity of $-$3.3$\pm$1.7~km\,s$^{-1}$ (see Section~3.2), we obtain a 
peculiar velocity of 22$^{+10}_{-6}$ km\,s$^{-1}$, which is much lower 
than required for a PN to belong to the halo.  
From these results we conclude that most probably Hu\,1-2 is not a halo PN.

\subsection{Excitation mechanism of the outer knots}

The H$\alpha$, [N~{\sc ii}] and [O~{\sc iii}] composite picture of 
Hu\,1-2 (Figure~\ref{color_img}) shows two bipolar knots at a distance 
$\sim$27\arcsec\ from the central source.  
The ionization and excitation conditions of these knots are studied 
in detail in this section using our narrow-band images and long-slit 
intermediate-dispersion spectroscopy.  
Since the SE knot is partially superimposed by a field star, our analysis 
will be mainly based on the NW knot, in which a bow-shock-like structure 
is otherwise clearly observed. 

\begin{figure}
\epsfig{file=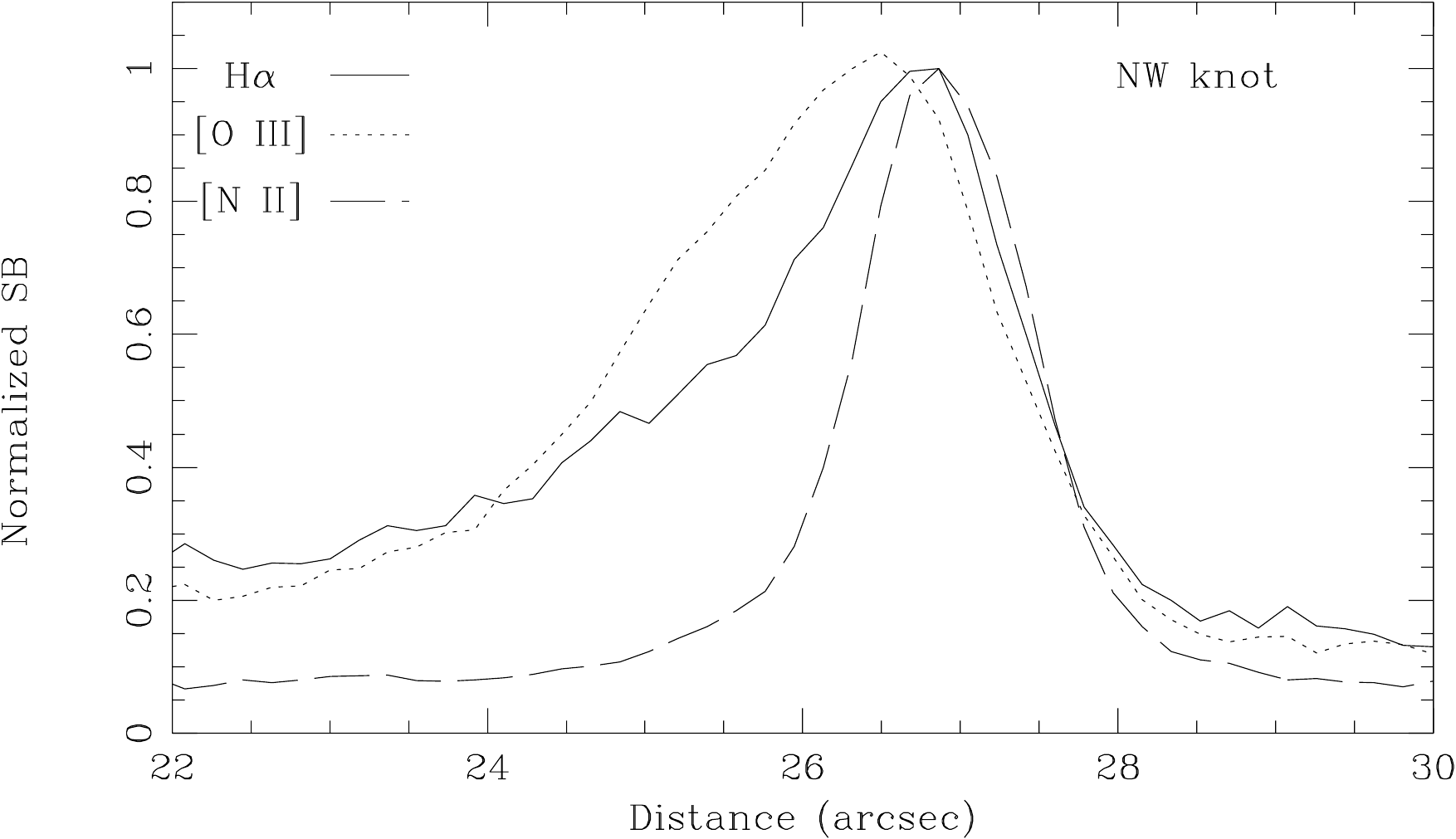,width=8cm,angle=0}
\caption{
Brightness profiles of the [O~{\sc iii}] (dotted line), H$\alpha$ (solid 
line), and [N~{\sc ii}] (dashed line) emission lines along the axis that 
connects the center of Hu~1-2 with the NW knot.  The profiles have been 
normalized to their intensity peaks for a fair comparison.  Distances are 
measured from the central star (in arcsec). }
\label{profile}
\end{figure}

Figure~\ref{profile} shows the NOT ALFOSC H$\alpha$, [O~{\sc iii}] 
$\lambda$5007, and [N~{\sc ii}] $\lambda$6583 surface brightness 
distributions (or 1-D profiles) projected along the line that 
connects the NW knot to the center of Hu\,1-2.  
The knot has single-peaked profiles in the three emission lines, 
but they show notable differences.  
The H$\alpha$ and [O~{\sc iii}] emissions extend inward from the head of 
the bow-shock toward the central source, whereas the distribution of the 
[N~{\sc ii}] line has a single peak.  
The positions of the peaks in different emission lines also differ;   
the [N~{\sc ii}] peak is the farthest from the central source, 
the H$\alpha$ peak is located at an intermediate distance, and 
the [O~{\sc iii}] peak is the closest to the central source.  
This same pattern is observed in the bow-shock structures 
of IC\,4634 and NGC\,7009 (Raga et al.\ \citealt{raga2008}).  
The position of the [N~{\sc ii}] peak coincides with that of 
the H$_2$ emission, further confirming the small thickness of 
the shell of low-ionization material.  
For a distance of 3.5~kpc (Miranda et al.\ \citealt{mir12a}), the spatial 
shifts between the H$\alpha$ and the [N~{\sc ii}] and [O~{\sc iii}] peaks 
are $\sim$3.5$\times$10$^{15}$~cm and $\sim$10$^{16}$~cm, respectively.  
These are similar, but somewhat larger, than the shifts of a few times 
10$^{15}$~cm observed in the outer bow shocks of IC\,4634 and NGC\,7009.

\begin{figure*}
\begin{center}
\epsfig{file=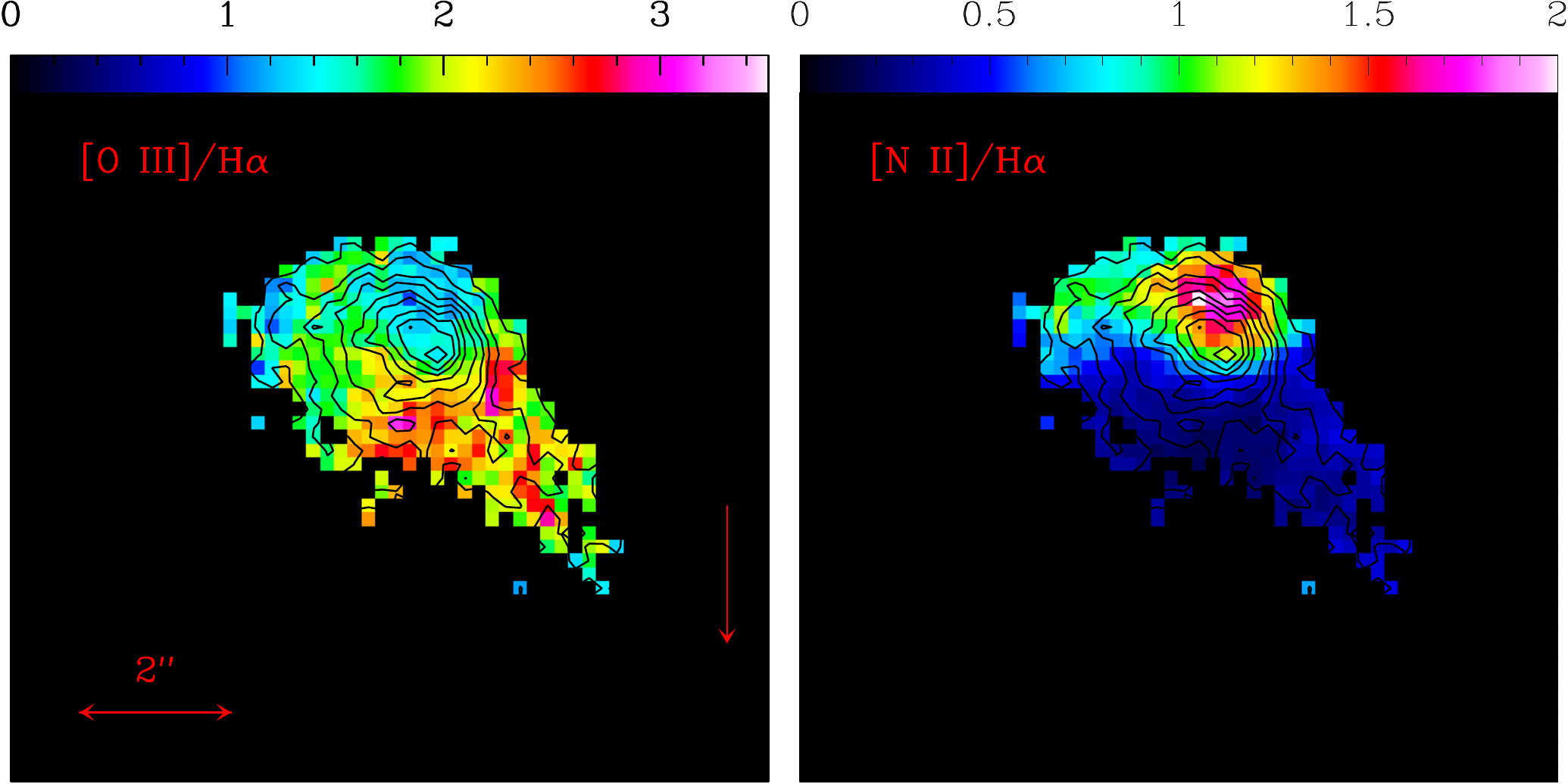,width=9.0cm,angle=-90}
\caption{
[O~{\sc iii}]/H$\alpha$ (left) and [N~{\sc ii}]/H$\alpha$ (right) ratio 
maps of the NW knot.  The black contours over-plotted on the ratio maps 
trace the intensity in the H$\alpha$ line. The red arrow in the left panel 
points to the direction of the central star of Hu\,1-2. }
\label{ratios}
\end{center}
\end{figure*}

Figure~\ref{ratios} shows the [O~{\sc iii}]/H$\alpha$ and [N~{\sc 
ii}]/H$\alpha$ ratio maps of the NW knot obtained at each pixel with 
surface brightness above a threshold value of 3~$\sigma$ from the 
background. 
Abrupt changes in these emission line ratios are observed at the position 
of the H$\alpha$ peak.  At the leading edge of the bow-shock, the 
[O~{\sc iii}]/H$\alpha$ has values of 1.2--1.8.  
Behind the H$\alpha$ peak, the [O~{\sc iii}]/H$\alpha$ ratio increases 
up to values 2--3, following the ionization stratification also found 
in the FLIERs of other PNe (Balick et al. \citealt{balick1998}; Riera 
\& Raga \citealt{rr2007}).  
At the head of the bow-shock, where the [O~{\sc iii}]/H$\alpha$ 
ratio declines, the [N~{\sc ii}] emission is enhanced, with 
[N~{\sc ii}]/H$\alpha$ ratios of 2--3, significantly raised from 
the value $\leq$0.5 detected upstream. 
In short, the NW knot of Hu\,1-2 shows a low-excitation [N~{\sc ii}]-bright 
head and a high-excitation [O~{\sc iii}]-bright, arc-shaped wing and 
upstream region.

Our intermediate-dispersion spectroscopy cannot resolve spatially the 
ionization structure seen in the NOT ALFOSC images, but the use of 
additional line diagnostic diagrams provides further constraint on the 
excitation conditions of this knot.  The spectrum of the NW knot is 
characterized by lines of intermediate-to-high excitation.  Besides the 
H~{\sc i} Balmer lines, the strongest optical lines are [O~{\sc ii}] 
$\lambda$3727, [O~{\sc iii}] $\lambda\lambda$4959,\,5007, [N~{\sc ii}] 
$\lambda\lambda$6548,\,6583, and He~{\sc ii} $\lambda$4686.  We also 
detect the strong, high excitation [Ne~{\sc iii}] $\lambda$3967 line in 
the NW knot.  The intensities of the [O~{\sc ii}] $\lambda$3727, 
[N~{\sc ii}] $\lambda\lambda$6548,\,6583 and [S~{\sc ii}] 
$\lambda\lambda$6716,\,6731 emission lines relative to H$\beta$ are much 
stronger in the NW knot than in the inner and outer regions.

\begin{figure*}
\begin{center}
\epsfig{file=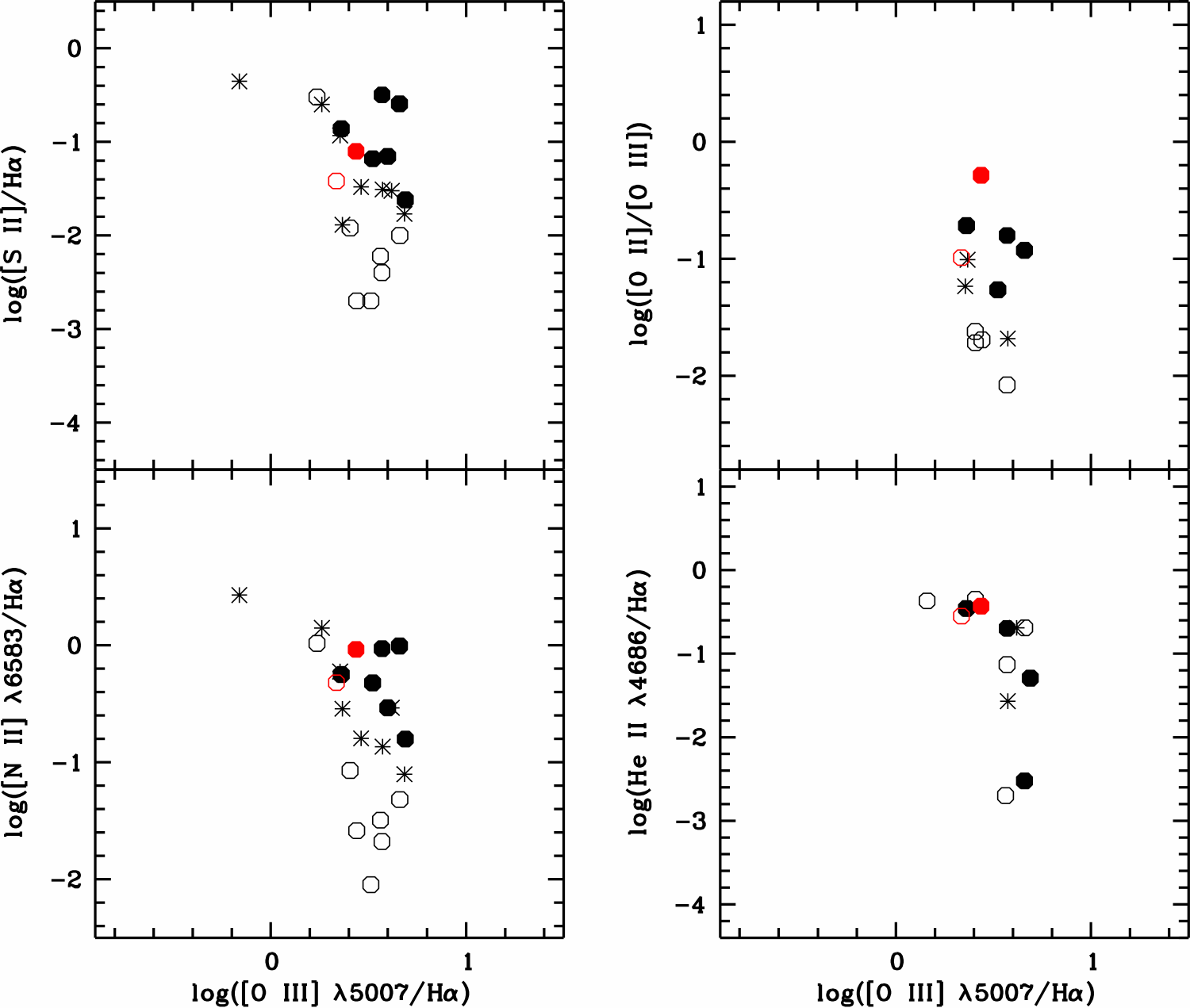,width=0.9\textwidth,angle=0}
\caption{
Diagnostic diagrams of several emission line ratios for rims/shells (open 
circles), low-ionization structures projected onto the nebular shells 
(stars), and outflows (filled circles) for a sample of PNe:  ETHOS\,1, 
He\,1-1, IC\,4634, K\,1-2, KjPn\,8, NGC\,6543, NGC\,6826, NGC\,7009, 
NGC\,7354, and NGC\,7662 (Balick et al.\ \citealt{balick1994}; Contreras 
et al.\ \citealt{con10}; Exter, Pollacco \& Bell \citealt{ext03}; 
Gon\c calves et al.\ \citealt{gon03,gon09}; Guerrero et al.\ 
\citealt{guerrero2008}; L\'opez-Mart\'\i n et al.\ \citealt{martin02}; 
Miszalski et al.\ \citealt{mis11a}). 
The emission line ratios for Hu\,1-2 are shown in red.  
}
\label{diag-obs}
\end{center}
\end{figure*}

Figure~\ref{diag-obs} shows the observed line ratios in different regions 
of Hu\,1-2 compared to those measured in the rims/shells and 
low-ionization structures of a sample of PNe using a set of diagnostic 
diagrams adopted from Raga et al.\ \cite{raga2008}.  
In these diagrams, the data points associated to the rims and shells of 
PNe trace the locus of photoionized gas.  The data points corresponding 
to low-ionization structures have been split into two different groups.  
The first one includes low-ionization structures that are projected inside 
the nebular shells (e.g., the FLIERs of NGC\,6826), whereas the second 
group consists of those projected outside the nebular shells (e.g., the 
bow-shock features of IC\,4634).  
This very simple classification is expected to assign mostly low-velocity 
photoionized structures to the first group, whereas the second group will 
be mostly populated by fast outflows, where irradiated shocks are 
expected to occur.  

In the [S~{\sc ii}]/H$\alpha$ and [N~{\sc ii}]/H$\alpha$ versus 
[O~{\sc iii}]/H$\alpha$ diagrams, the data points of Hu\,1-2 are located 
in the regions occupied by other PNe (Figure~\ref{diag-obs}).  In 
particular, the emission line ratios for the NW knot are compatible with 
those of the low-ionization structures in other PNe, with the [S~{\sc 
ii}]/H$\alpha$ and [N~{\sc ii}]/H$\alpha$ ratios consistently higher (by 
definition) than those derived for the photoionized rims and shells of 
PNe.  Furthermore, the [O~{\sc iii}]/H$\alpha$ ratios of outflows tend 
to be higher than those of low-ionization structures projected inside 
the nebular shells.  This can be evidence of an additional source of 
excitation, i.e., the shocks, particularly at larger distances from the 
central source, where the ionizing flux of photons is reduced and thus 
lower [O~{\sc iii}]/H$\alpha$ ratios would have been expected instead.  
It is also interesting to note that the NW knot of Hu\,1-2 shows higher 
[O~{\sc ii}]/[O~{\sc iii}] and He~{\sc ii}/H$\alpha$ ratios than those 
typically seen in the low-ionization structures and outflows of other PNe. 
The He~{\sc ii}/H$\alpha$ ratios in the NW knot is also compatible with 
that observed in the inner region of Hu\,1-2 (Figure~\ref{diag-obs}, 
bottom right).  These patterns observed in Hu\,1-2 might be a consequence 
of the very high temperature of its central star (125\,000~K; Hyung, 
Pottasch \& Feibelman \citealt{hyu04}).  The NW knot of Hu\,1-2 probably 
harbors a significant amount of high-excitation material, and its relative 
amount of the high- to low-excitation material could be higher than that 
in the outflows and low-ionization structures of other PNe, making it a 
very peculiar structure.

\begin{figure*}
\begin{center}
\epsfig{file=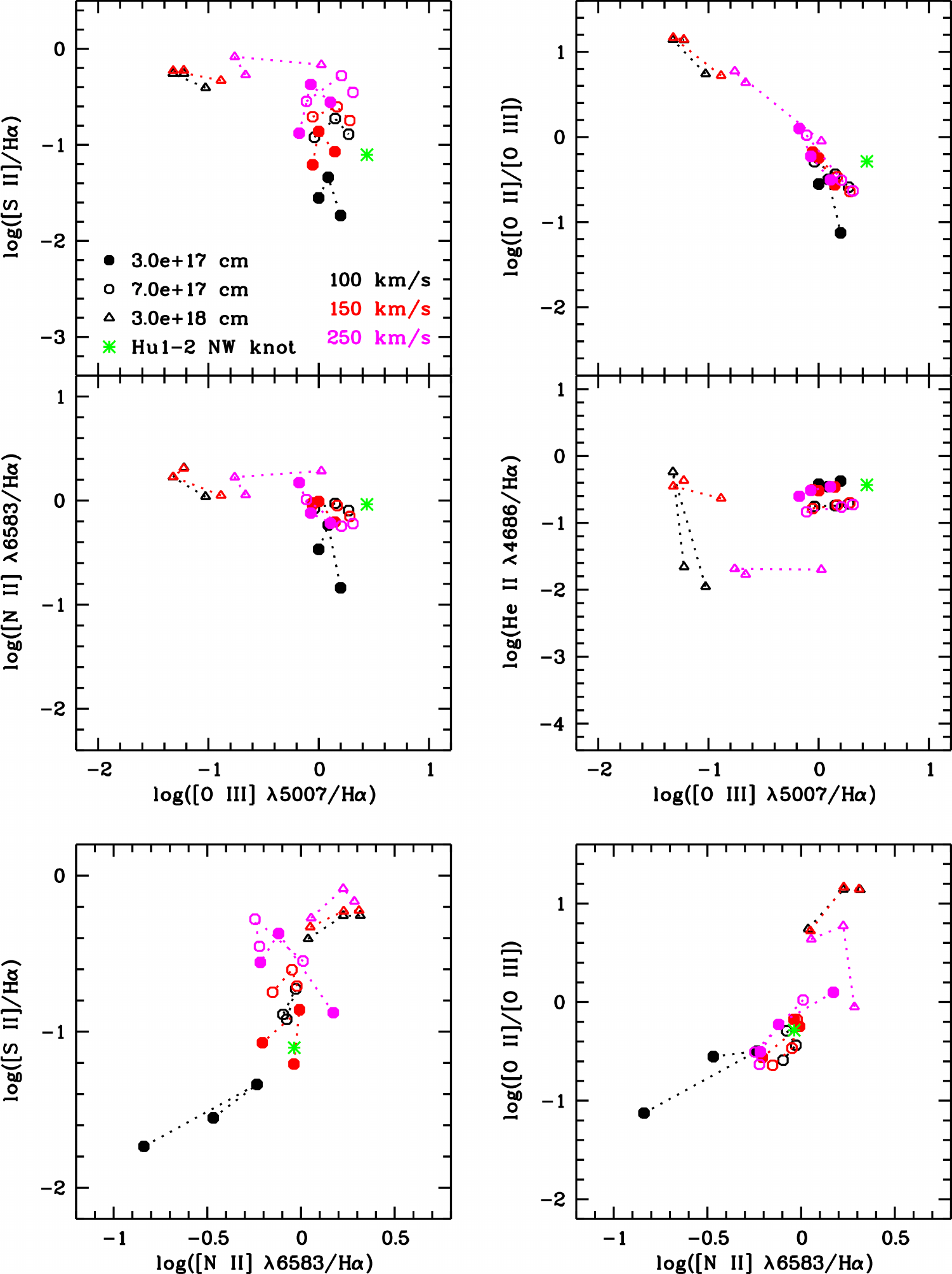,width=0.9\textwidth,angle=0}
\caption{
Similar to Figure~\ref{diag-obs}, but comparing the emission line ratios 
predicted for a shocked cloudlet moving away from a photoionizing source 
with those seen in the NW knot of Hu\,1-2 (the green star) for different 
distances from the central star and at different expansion velocities 
(see description of the symbols in the upper-left panel).  Symbols of the 
same models are connected with dotted lines to show the time-evolution 
sequences. }
\label{diag-model}
\end{center}
\end{figure*}

The observed line ratios in the NW knot of Hu\,1-2 are also compared 
in Figure~\ref{diag-model} with the predictions for shocked regions in 
high-velocity knots of PNe developed by Raga et al.\ \cite{raga2008}. 
This comparison helps to diagnose the excitation mechanisms 
(photoionization, shocks, or both) in the NW knot. 
These models assume a cloudlet traveling away from a photoionizing source 
into a uniform medium ($\sim$10$^{2}$~cm$^{-3}$).  The simulations include 
transfer of the ionizing radiation and a non-equilibrium ionization 
network of many ionic species.  In Figure~\ref{diag-model}, We present 
the predicted line ratios from an ``ad hoc'' model for Hu\,1-2 where a 
high-density ($\sim$10$^{3}$~cm$^{-3}$), high-velocity ($v_{\rm c}$ = 
100, 150 and 250~km\,s$^{-1}$) cloudlet with the chemical abundances 
measured in Hu\,1-2 is moving away from a 100\,000-K central star. 
Cases with different distances from the central source 
(3$\times$10$^{17}$, 7$\times$10$^{17}$ and 3$\times$10$^{18}$~cm) have 
been simulated.  Those are axisymmetric simulations, and the initial 
temperature of the clump/cloudlet was set to be 10$^{4}$~K.  In order 
to compare the observations and the numerical simulations, we integrated 
the computed emission line coefficients over the entire emitting volume 
to derive the emission line luminosities of the NW knot of Hu\,1-2. 
Predicted line ratios for different integration times since the start 
of the shock are summarized in Table~\ref{models}.

The predicted [S~{\sc ii}]/H$\alpha$, [N~{\sc ii}]/H$\alpha$, 
[O~{\sc ii}]/[O~{\sc iii}] and He~{\sc ii}/H$\alpha$ line ratios 
in Table~\ref{models} show a large scatter, but they are 
generally comparable to those derived from the observations of 
Hu\,1-2.  
The plots in Figure~\ref{diag-model} show that the best models to reproduce 
the observed line ratios in the NW knot of Hu\,1-2 are those for a knot at 
distances of a few times 10$^{17}$ cm.  
Models at the largest distance, 3$\times$10$^{18}$ cm, generally 
predict too low [O~{\sc iii}]/H$\alpha$ ratios, which can be 
explained by the greater dilution of the ionizing flux from the 
central star.  
For models where the knot has smaller distances ($\sim$10$^{17}$~cm) 
from the central source (this is equivalent to changes in the ionizing 
flux at the clump), 
all predicted line ratios agree with those observed in the NW knot of 
Hu\,1-2 within a factor of 2 to 3. 
Given the simplicity and limitations of our models (a spherical clump 
moving away from the central source through a constant medium, where 
the absorption of the photons by nebular gas between the star and the 
clump is not accounted for), the discrepancies between the observed line 
ratios and those predicted can be considered to be reasonable.


The plots in Figure~\ref{diag-model} also help to constrain the knot 
velocity.  Models at low velocity (100~km\,s$^{-1}$) tend to predict 
[S~{\sc ii}]/H$\alpha$, [N~{\sc ii}]/H$\alpha$, and [O~{\sc ii}]/[O~{\sc 
iii}] line ratios lower than those observed in Hu\,1-2.  On the contrary, 
models at high velocity (250 km~s$^{-1}$) predict too high values for these 
line ratios.  Therefore, the knot velocity seems constrained in the range 
100--250~km\,s$^{-1}$.  We note that this knot velocity is smaller than 
the outflow velocity $>$340~km\,s$^{-1}$ derived from a spatio-kinematical 
model of the outflow (Miranda et al.\ \citealt{mir12a}).

Finally, we remark the persistent difficulty of our models to reproduce 
the observed [O~{\sc iii}]/H$\alpha$ line ratios.  This issue can be 
alleviated by increasing the oxygen abundance assumed by our model.  We 
tested this by increasing the oxygen abundance, and found that increasing 
the abundance within the uncertainty of our abundance determination 
produced some improvement (not shown in Figure~\ref{diag-model}); however 
it was insufficient to reproduce the observed [O~{\sc iii}]/H$\alpha$ line 
ratio.

\begin{table*}
\begin{minipage}{110mm}
\caption{Shock model results compared with observations of the Hu\,1-2 NW 
knot.}
\label{models}
\centering
\begin{tabular}{lccccccc}
\hline
\multicolumn{1}{l}{$v_{\rm c}$}  &
\multicolumn{1}{l}{Distance} &
\multicolumn{1}{c}{$t$}     &
\multicolumn{1}{c}{[O~{\sc ii}]/H$\alpha$} &
\multicolumn{1}{c}{[O~{\sc iii}]/H$\alpha$} &
\multicolumn{1}{c}{[N~{\sc ii}]/H$\alpha$} &
\multicolumn{1}{c}{[S~{\sc ii}]/H$\alpha$} &
\multicolumn{1}{c}{He~{\sc ii}/H$\alpha$} \\
\multicolumn{1}{l}{(km\,s$^{-1}$)} &
\multicolumn{1}{l}{($\times$10$^{18}$~cm)} &
\multicolumn{1}{c}{(years)} &
\multicolumn{1}{c}{} &
\multicolumn{1}{c}{} &
\multicolumn{1}{c}{} &
\multicolumn{1}{c}{} &
\multicolumn{1}{c}{} \\
\hline
100 & 0.3 &  50 & 0.280 & 1.000 & 0.340 & 0.028 & 0.380 \\
    &     & 200 & 0.388 & 1.220 & 0.582 & 0.046 & 0.347 \\
    &     & 350 & 0.118 & 1.580 & 0.145 & 0.018 & 0.421 \\
\\
100 & 0.7 &  50 & 0.468 & 0.915 & 0.839 & 0.120 & 0.176 \\
    &     & 200 & 0.515 & 1.410 & 0.938 & 0.188 & 0.181 \\
    &     & 350 & 0.482 & 1.870 & 0.802 & 0.129 & 0.193 \\
\\
100 & 3.0 &  50 & 0.519 & 0.094 & 1.090 & 0.393 & 0.011 \\
    &     & 200 & 0.659 & 0.048 & 1.680 & 0.553 & 0.578 \\
    &     & 350 & 0.826 & 0.060 & 2.070 & 0.554 & 0.022 \\
\\
150 & 0.3 &  50 & 0.588 & 0.882 & 0.916 & 0.062 & 0.313 \\
    &     & 200 & 0.566 & 1.000 & 0.981 & 0.138 & 0.302 \\
    &     & 350 & 0.384 & 1.400 & 0.622 & 0.085 & 0.343 \\
\\
150 & 0.7 &  50 & 0.588 & 0.882 & 0.951 & 0.196 & 0.164 \\
    &     & 200 & 0.501 & 1.470 & 0.894 & 0.249 & 0.185 \\
    &     & 350 & 0.439 & 1.920 & 0.705 & 0.179 & 0.200 \\
\\
150 & 3.0 &  50 & 0.684 & 0.130 & 1.120 & 0.468 & 0.233 \\
    &     & 200 & 0.704 & 0.048 & 1.690 & 0.592 & 0.352 \\
    &     & 350 & 0.835 & 0.060 & 2.040 & 0.597 & 0.429 \\
\\
250 & 0.3 &  40 & 0.838 & 0.666 & 1.486 & 0.132 & 0.250 \\
    &     & 120 & 0.503 & 0.847 & 0.759 & 0.425 & 0.306 \\
    &     & 200 & 0.401 & 1.276 & 0.606 & 0.278 & 0.346 \\
\\
250 & 0.7 &  40 & 0.815 & 0.775 & 1.025 & 0.283 & 0.146 \\
    &     & 120 & 0.500 & 1.606 & 0.568 & 0.525 & 0.173 \\
    &     & 200 & 0.477 & 2.047 & 0.600 & 0.352 & 0.188 \\
\\
250 & 3.0 &  40 & 0.943 & 0.217 & 1.132 & 0.535 & 0.017 \\
    &     & 120 & 1.026 & 0.173 & 1.679 & 0.826 & 0.021 \\
    &     & 200 & 0.946 & 1.056 & 1.925 & 0.683 & 0.020 \\
\\
\multicolumn{3}{l}{Observations$^{a}$} & 1.415 & 2.735 & 0.920 & 0.080 & 0.368\\
\hline
\end{tabular}
\begin{description}
\item[$^{a}$] Observations of the NW knot of Hu\,1-2.
\end{description}
\end{minipage}
\end{table*}

\subsection{The jet-launching engine of Hu\,1-2}

The total mass, momentum and mechanical luminosity of the jets can be 
used to assess the jet-launching engine of a PN or proto-PN (e.g., 
Bujarrabal et al.\ \citealt{buj01}; Blackman \& Lucchini \citealt{bl14}).  
Similar diagnosis is carried out for Hu\,1-2 below.  
The ionized mass of the NW knot can be estimated from its H$\beta$ flux.  
Using the total H$\alpha$ flux of the NW knot derived from the NOT H$\alpha$ 
flux-calibrated image presented in Figure~\ref{grey_img} and the H$\alpha$ to 
H$\beta$ ratio derived 1-D spectrum in Table~2, we derived a total H$\beta$ 
flux of $\sim$3$\times$10$^{-15}$ erg\,cm$^{-2}$\,s$^{-1}$ for the NW knot.  
Adopting a distance of 3.5~kpc (Miranda et al. \citealt{mir12a}), and 
an angular radius of 2\arcsec, an electron temperature of 10$^{4}$~K, 
and a filling factor of 0.4, we estimated an electron density of 
$\sim$200~cm$^{-3}$ and an ionized mass of $\sim$2$\times$10$^{-5}$ 
$M_{\sun}$ for the bipolar knots of Hu\,1-2.

The dynamical age of the bipolar knots of Hu\,1-2 is well constrained 
to be 1375~yr under the assumption that the knots have been moving 
ballistically since their ejection (Miranda et al.\ \citealt{mir12a}). 
Under the same assumption, the ejection time-scale of these knots can 
be estimated from the ratio between their angular size, $\sim$4\arcsec, 
and their distance to the central star, $\sim$27\arcsec.  
This result in an ejection time-scale ($\tau_{\rm jet}$) $\sim$200~yr.  
The averaged mass-loss rate during the ejection would 
then be $\sim$1$\times$10$^{-7}$ $M_{\sun}$~yr$^{-1}$.

The mechanical luminosity of the jets can be estimated using the equation 
\begin{equation}
\label{eq3}
L_{\rm mec} = \frac{\frac{1}{2} M_{\rm knot} v_{\rm exp}^{2}}{\tau_{\rm jet}}, 
\end{equation}
where $M_{\rm knot}$ is the total ionized mass of the knots, 
$v_{\rm exp}$ is the expansion velocity (340~km\,s$^{-1}$; Miranda et 
al.\ \citealt{mir12a}), and $\tau_{\rm jet}$ is the jet ejection 
time-scale ($\sim$200~yr).  
Thus the jets mechanical luminosity of Hu\,1-2 would be $\sim$0.5~$L_{\sun}$.  
The total energy of the jets is $\sim$1.2$\times$10$^{43}$~erg, which is 
lower than the kinetic energies of fast outflows in proto-PNe (e.g., the 
sample studied by Bujarrabal et al.\ \citealt{buj01}) by at least one 
order of magnitude.  
The momentum estimated for the bipolar knots of Hu\,1-2 is 
$\sim$1.4$\times$10$^{36}$~g\,cm\,s$^{-1}$, also much lower than those 
(10$^{37}$--10$^{40}$~g\,cm\,s$^{-1}$) observed in proto-PNe, whose high 
linear momenta defy any easy interpretation (Bujarrabal et al.\ 
\citealt{buj01}).

\begin{table}
\begin{minipage}{60mm}
\centering
\caption{Physical properties of collimated outflows in PNe. 
}
\label{luminosity}
\begin{tabular}{lcccc}
\hline
PN & $M_{\rm knot}$ & $v_{\rm exp}$  & $\tau_{\rm jet}$ & $L_{\rm mec}$\\
   & ($M_{\sun}$)   & (km\,s$^{-1}$) & (yr)             & ($L_{\sun}$) \\
\hline
Hu\,1-2            & $\sim$2$\times$10$^{-5}$ & 340           & 200        & 0.5      \\
The Necklace$^{a}$ & $\sim$10$^{-3}$          & 95(N), 115(S) & 3700--8000 & 0.1--0.23\\
NGC\,6778$^{a}$    & 1.5$\times$10$^{-3}$     & 270, 460      & 1700       & 5.3      \\
\hline
\end{tabular}
\begin{description}
\item[$^{a}$] From Tocknell, De~Marco \& Wardle \cite{toc14}. 
\end{description}
\end{minipage}
\end{table}

The parameters associated to the jet launching in Hu\,1-2 and those 
studied by Tocknell, De~Marco \& Wardle \cite{toc14} are presented 
in Table~\ref{luminosity}.  
A comparison among them indicates that the bipolar knots of Hu\,1-2 
are lighter, less massive than those of the Necklace Nebula and 
NGC\,6778, whereas their ejection time-scale is much shorter.  The 
two effects combined result in very similar mechanical luminosities 
for all sources in Table~\ref{luminosity}.  We note the large 
uncertainty in the estimate of the ejection time-scale of the knots 
of Hu\,1-2.  Dynamical effects are certainly playing a crucial role 
in the evolution of the outflow, compressing the head of the flow 
(reducing the apparent ejection time-scale) or lagging material 
upstream (increasing the apparent ejection time-scale).  Thus this 
ejection time-scale could be quite uncertain and so is the derived 
jets mechanical luminosity ($L_{\rm mec}$) of Hu\,1-2.

It is interesting to note that the jets mechanical luminosity of 
NGC\,6778 is extremely high (Table~\ref{luminosity}), if we adopt the 
parameters given in Tocknell, De~Marco \& Wardle \cite{toc14}.  This 
PN, which has a similar distance ($\sim$2.6~kpc; Tocknell, De~Marco \& 
Wardle \citealt{toc14}) as Hu\,1-2, has a highly disrupted equatorial 
ring and a binary central star (Miszalski et al.\ \citealt{mis11b}; 
Guerrero \& Miranda \citealt{gm12}).  
Both NGC\,6778 and the Necklace Nebula in Table~\ref{luminosity} have 
a post-common envelope (post-CE) close binary in the center (Miszalski 
et al.\ \citealt{mis11b}; Corradi et al.\ \citealt{corradi11}).  Like 
NGC\,6778, Hu\,1-2 displays an overall bipolar structure and a likely 
disrupted equatorial ring.  This resemblance suggests that Hu\,1-2 
might have similar jet-launching mechanisms as the other two PNe. 

\section{Conclusions}

We have presented a thorough imaging and spectroscopic study of the main 
nebular regions and bipolar outflows of Hu\,1-2.  The physical structure 
of the main nebula can be described as bipolar, although its velocity 
field cannot be described by a simple hour-glass law.  The inclination 
of the bipolar lobes is $\leq$10\degr\ and its polar expansion velocity 
$\simeq$150~km\,s$^{-1}$, resulting in a lower-limit expansion age 
$\lesssim$1100~yr.  Neither the morphology nor the kinematics of the 
central, $z$-shaped innermost region of Hu\,1-2 can be simply described 
as a waist or torus located between the two bipolar lobes.  The knotty 
morphology of this region and its relatively high expansion velocity, 
$\simeq$40~km\,s$^{-1}$, are suggestive of a ``broken'' equatorial ring.  
The distorted velocity field of the bipolar lobes and the complex morphology 
of the equatorial regions make us conclude that Hu\,1-2 has experienced 
notable violent dynamical processes during its formation.  Similar 
interpretation has been given to the disrupted equatorial regions of a few 
bipolar PNe known to harbor binary systems (e.g., NGC\,6778, NGC\,7354; 
Contreras et al. \citealt{con10}; Miszalski et al. \citealt{mis11b}; 
Guerrero \& Miranda \citealt{gm12}).  
It is tempting to conclude that the fast bipolar knots present in 
all these sources are dynamical agents which caused the disrupted 
equatorial features.

It is interesting to remark here the apparent discrepancies implied by the 
chemical abundances of Hu\,1-2.  The high He/H and N/O abundance ratios 
suggest a Type~I nature for Hu\,1-2, which indicates that it descended from 
a relatively massive intermediate-mass progenitor star 
($\gtrsim$4~$M_{\sun}$; e.g., Karakas et al.\ \citealt{kar09}). 
On the other hand, the low abundances of 
the heavy elements of Hu\,1-2 suggest it formed long time ago, probably in 
an early-stage of the chemical evolution of the Galaxy, and thus it 
corresponds to a low-mass progenitor. 
Cases of Type~I PNe with very low abundances of $\alpha$ 
elements can be interpreted as a result of formation of a massive PN 
progenitor from unmixed (i.e., metal-poor) interstellar material (Milingo 
et al.\ \citealt{mil10}).  Therefore, the chemical composition of Hu\,1-2 
probably does not represent that of its current environment.  This 
might indeed be the case.  At a galactocentric distance $\sim$9\,kpc, as 
estimated from the distance to Hu\,1-2 ($\sim$3.5\,kpc; Miranda et al.\ 
\citealt{mir12a}), its Galactic longitude (86\fdg5) and the galactocentric 
distance of the Sun (8.3\,kpc; Gillessen et al.\ \citealt{gil09}), the 
oxygen abundance of the interstellar medium is $\sim$4$\times$10$^{-4}$ 
(Henry et al.\ \citealt{hen10}, and Figure~1 therein), which is much 
higher than that of Hu\,1-2 (see Table~\ref{elemental}).  Alternatively, 
Type~I PNe have long been connected to bipolar morphology (e.g., Peimbert 
\& Torres-Peimbert \citealt{ptp83}) which is likely the result of 
binary interactions (e.g., Soker \citealt{sok97}) or common 
envelope binary interaction (e.g., Zijlstra \citealt{zijl07}; 
De~Marco \citealt{marco09}; Miszalski et al. \citealt{mis09}). 
The abundance pattern of Hu\,1-2 might be due to the effects 
of binary interactions on the evolution of the progenitor and the 
composition of the subsequent PN, although so far it is not clear 
whether there is a binary central star in Hu\,1-2.  Detailed discussion 
of this point is beyond the scope of this paper. 
We also discussed the possibility of Hu\,1-2 being a halo PN, given 
that it is quite out of the Galactic plane ($b$ = $-$8\fdg8).  Although 
the oxygen abundances of Hu\,1-2 and its distance to the Galactic plane 
marginally support a halo nature, its low peculiar radial velocity 
obviously argues against a halo nature for Hu\,1-2.

The collimated bipolar outflows of Hu\,1-2 are particularly interesting.  
They show a notable bow-shock-like morphology.  
The emission within these structures is found to be highly stratified, 
with [N~{\sc ii}] peaking at the leading edge of the bow-shock and 
[O~{\sc iii}] mostly occupying the star-facing region.  
This profile is similar to that found in the bow-shock structures of 
PNe such as IC\,4634 and NGC\,7009.  
Previous spatio-kinematical studies have suggested an expansion velocity  
$>$340~km\,s$^{-1}$ for the outer knots of Hu\,1-2.  
This high expansion velocity of the bipolar knots, together with the 
bow-shock-like morphology associated with them, indicate that the 
high-velocity, collimated bipolar outflows are moving through the 
interstellar medium like ``bullets''.

The line ratios of the NW knot of Hu\,1-2 and those from the collimated 
outflows of other PNe are generally consistent with those of other 
low-ionization structures, but they exhibit higher [O~{\sc iii}]/H$\alpha$ 
ratios.  This line ratio is particularly high in Hu\,1-2, indicating that 
both low- and high-excitation gas are present at the location of this 
feature with significant amounts.  The observed line ratios of the NW 
knot of Hu\,1-2 have been modeled using ``ad~hoc'' simulations of a fast 
moving cloudlet in a medium with homogeneous density.  
The model predictions are generally consistent with the observed line 
ratios, thus confirming that the excitation of the bipolar knots of 
Hu\,1-2 can be explained by a mix of the UV radiation and shocks.  The 
models favor distance, nebular abundances, and stellar parameters 
consistent with those derived for Hu\,1-2.  As reported for the 
collimated outflows of other PNe and proto-PNe (e.g., Hen\,3-1475; 
Riera et al.\ \citealt{rie06}), the best knot velocity (100--250 
km\,s$^{-1}$) inferred from the models is below that derived by 
spatio-kinematical studies ($>$340 km~s$^{-1}$).  


Finally, we estimated the ionized mass of the NW knot of Hu\,1-2, and 
calculated the jets mechanical luminosity using the knots' mass and 
the ejection time-scale based on the dynamical age of the bipolar 
outflow.  The jets mechanical luminosity and the mass-loss rate of 
Hu\,1-2 generally agree, within the uncertainties, with those of the 
Necklace and NGC\,6778, both of which have a post-common envelope 
close binary central star.  The resemblance of Hu\,1-2 with NGC\,6778 
in nebular structure and morphology hints at a possibility that the 
former might have similar jet-launching engine as the latter. 

\section*{Acknowledgements}
XF and MAG acknowledge support from grant AYA 2011-29754-C03-02.  LFM is 
partially supported by grant AYA2011-30228-C3-01 (co-funded by FEDER funds) 
of the Spanish MICINN. 
AR is supported by grant MINECO AYA2011-30228-C03 (Spain). PFV and ACR are 
supported by CONACyT grant 167611 and DGAPA-PAPIIT (UNAM) grant IG100214.  
This paper is based on observations made with the 1.5m telescope operated 
by the Instituto de Astrof\'\i sica de Andaluc\'\i a at the Observatorio 
de Sierra Nevada, Granada, Spain, and with the Nordic Optical Telescope 
(NOT) and the Italian Telescopio Nazionale Galileo (TNG) operated at the 
Observatorio del Roque de los Muchachos, La Palma, Spain, by the Nordic 
Optical Telescope Scientific Association and the Fundaci\'on Galileo 
Galilei of the INAF (Istituto Nazionale di Astrofisica), respectively.
We thank Orsola De~Marco, the referee of this article, for her 
insightful comments which have greatly improved the quality of this 
article.


\end{document}